\documentclass[a4paper, 12pt]{article}
\usepackage[utf8]{inputenc}
\usepackage[T2A]{fontenc}
\usepackage[english]{babel}
\usepackage{amssymb}
\usepackage{amsmath}
\usepackage{textcomp}
\usepackage{indentfirst}
\usepackage{float}

\usepackage{units}
\usepackage{amsfonts}
\usepackage{graphicx} 
\usepackage{amstext} 
\usepackage{wasysym} 

\usepackage{epstopdf}
\usepackage{xcolor}
\usepackage{hyperref}
 
\definecolor{linkcolor}{HTML}{000000} 
\definecolor{urlcolor}{HTML}{000000} 
 
\hypersetup{pdfstartview=FitH,  citecolor=black, linkcolor=linkcolor,urlcolor=urlcolor, colorlinks=true}
\usepackage[left=2.5cm,right=2cm, top=2cm,bottom=2cm,bindingoffset=0cm]{geometry} 

\DeclareMathOperator{\diag}{diag}

\makeatletter
\newcommand{\l@abcd}[2]{\hbox to\textwidth{#1\dotfill #2}}
\renewcommand{\l@section}{\@dottedtocline{1}{2.5em}{2.3em}}
\renewcommand{\l@part}{\@dottedtocline{1}{1.5em}{2.3em}}
\renewcommand{\l@subsection}{\@dottedtocline{1}{3.5em}{2.3em}}
\makeatother

\usepackage[sorting=none,backend=bibtex,citestyle=numeric]{biblatex}
\bibliography{ising-bib.bib}

\usepackage{authblk}

\begin{document}

\title{ The $k$-flip Ising game}
\date{}
\author{Aleksandr Kovalenko$^{(a)}$, Andrey Leonidov$^{(a,b)}$}
\affil{{\small 
(a) P.N. Lebedev Physical Institute, Moscow, Russia\\
(b) Moscow Institute of Physics and Technology, Dolgoprudny, Russia\\
 }}

\maketitle

\abstract{The game-theoretic view on the classical Ising model – the noisy binary choice Ising game, - is a well-known conceptual framework that, on the one hand, highlights similar patterns between social or artificial agent-based systems and ensembles of interacting physical spins and, on the other hand, helps to stress important differences between the ones. The study considers one of such differences – the possibility of simultaneous decision making of several agents that is typical for game theoretic interaction. To do this we analyse a partially parallel discrete time dynamics of an Ising-type system of $N$ interacting binary units on a complete graph in which at each time steps $k$ arbitrarily chosen units can change their states is analysed. The particular problem under study is a $k$ - dependence of the decay of a metastable configuration into a stable one. The analysis is based on an explicit analytical calculation, for arbitrary noise, of the transition matrix characterising the $k$ - flip evolution as well as the first two moments of the distribution of the fraction of players choosing one of the two available strategies. First two moments of the first hitting time distribution for sample trajectories corresponding to transition from a metastable and unstable states to a stable one are considered. A nontrivial dependence of these moments on $k$ for the decay of a metastable state is discussed. A presence of the minima at certain $k_{\rm min}$ is attributed to a competition between $k$-dependent diffusion and restoring forces. }

\section*{\label{sec:level1}Introduction}

Studying the dynamics of Ising-type systems composed of interacting binary units is a focal point of research in many fields including, in particular,  
statistical physics \cite{glauber1963time,derrida1989dynamical,salinas2001introduction,cirillo2003metastability,neri2009cavity,avni2025nonreciprocal,bagnoli2025metastability,garces2025phase},  neural networks  \cite{little1974existence,shaw1974persistent,little1975statistical,hopfield1982neural,peretto1984collective,bastolla1998relaxation} and multi-agent systems \cite{blume2003equilibrium,bouchaud2013crises,antonov2021self,leonidov2022strategic,antonov2023transition,Leonidov2024,leonidov2024strategic,garnier2024unlearnable}. A striking feature of this research is that within given assumptions on noise, interaction with neighbours and external influence the basic ingredient of the dynamical description, the expression for the flip rate, and therefore the characteristic dynamical regimes, are the same in all the three domains.  

Let us note that there is a deep distinction between the characteristics dynamics of the Ising-type models with symmetric(reciprocable) \cite{glauber1963time,little1974existence,shaw1974persistent,little1975statistical,hopfield1982neural,peretto1984collective,salinas2001introduction,blume2003equilibrium,bouchaud2013crises,antonov2021self,antonov2023transition,Leonidov2024} and asymmetric (nonreciprocable) \cite{derrida1989dynamical,bastolla1998relaxation,garnier2024unlearnable,avni2025nonreciprocal,garces2025phase} interactions. In the present work we restrict our discussion to the case of symmetric interactions. 

One of the crucial features of describing the temporal evolution of Ising-type systems in discrete time is a specification of a number of units $k$ that can change their state at a given time moment. The corresponding alternatives are single-flip, partially parallel and parallel dynamics where, in a system of N units, for which $k=1$, $1 < k < N$ and $k=N$ respectively.  

In studies of dynamical evolution of spin systems in statistical physics the standard assumption is a single-flip dynamics introduced by Glauber \cite{glauber1963time}. It is indeed a reasonable assumption because in realistic description of spin systems kinetics simultaneous flipping of spin clusters is naturally considered as having low probability. Partially parallel dynamics is most often discussed in the context of accelerating the Monte-Carlo simulations \cite{swendsen1987nonuniversal,niedermayer1988general,wolff1989collective} are used to accelerate Monte Carlo simulations. In particular, the Swendesen-Wang algorithm \cite{swendsen1987nonuniversal} is based on using the Fortuin-Kasteleyn cluster decomposition \cite{kasteleyn1969phase,fortuin1972random} developed for efficient description of the structure of partition function in the Potts model. Partially parallel/diluted dynamics for the two-dimensional Ising model was considered in the recent paper \cite{bagnoli2025metastability}. Two-spin parallel dynamics was also considered in the framework of dynamical message passing approach to studying dynamical properties of Ising model on graphs in \cite{neri2009cavity}.  Parallel dynamics was studied in \cite{derrida1989dynamical} in a general setting and in \cite{cirillo2003metastability} for the two-dimensional Ising model. 

The choice between sequential and parallel dynamics plays an important role in construction of models of (associative) memory based on  neural networks. In particular, the papers \cite{little1974existence,little1975statistical,little1978analytic} analyse the possible role of parallel dynamics in reaching long-lived states of a neural network, while in \cite{shaw1974persistent} it was assumed that at each time step a Poisson-distributed random number of neurons could switch their states. Finally, in the foundational paper \cite{hopfield1982neural} a Glauber-type sequential dynamics was assumed. The models discussed in all of these papers have close parallels with studies of spin system dynamics and Ising games on graphs. 

In game theory simultaneous decisions of agents are rather a norm than an exception. In this context a single-flip assumption in describing dynamical properties of noisy discrete choice games looks somewhat unnatural and simply borrowed from Glauber dynamics  \cite{glauber1963time,salinas2001introduction}. The dynamics of Ising games on graphs was previously studied in the single-flip approximation using an explicit Markov chain type description of system trajectory \cite{blume2003equilibrium} as well as by considering the corresponding Langevin \cite{bouchaud2013crises} and Fokker-Planck \cite{Leonidov2024} dynamics. In all these cases the arising description of a game was given in terms of a "free utility" (an analogue of the free energy in statistical physics) dependent, in the case of the Ising game on complete graph, on a single collective variable and combining the myopic utility of current configuration and its entropy.  Partially parallel dynamics was considered in the mean field analysis of strategic stiffening/cooling \cite{leonidov2024strategic}.

An important generic feature of Ising-type systems is a presence of long-lived metastable states. For the single flip dynamics their decay into stable configurations is exponentially long which a very important feature of modelling noisy binary choice decisions in socioeconomic models \cite{bouchaud2013crises}, see also  \cite{antonov2021self,antonov2023transition} for an analysis of acceleration of this decay through a self-excitation mechanism in a noisy binary choice (Ising game). It is of clear interest to study the properties of the decay of metastable states in the Ising-type on the type of dynamics (single-flip, partially or fully parallel). In particular, it is interesting to check whether the naive expectation of accelerating this decay by increasing the number of simultaneously evolving units.  

In the present study we focus on analysing the $k$ - dependence of the decay time of metastable states at an example of a noisy binary choice (Ising) game \footnote{ Let us stress that the results obtained are, within taken assumptions, are also valid for dynamics of spin systems and neural networks. } allowing for simultaneous decisions of a randomly chosen group of agents. In this paper we consider a simplest setting of a noisy binary choice (Ising) game of $N$ agents located in the vertices of a complete graph played in discrete time in which at each time moment $k$ randomly chosen agents, $k \in [1,N]$, are allowed to reconsider their current strategies. As has been already mentioned, it is to be expected that a transition from single-flip to "multi-flip" dynamics involving randomly formed clusters of $k$ agents should directly affect evolution pace by accelerating it. A quantitative assessment of this acceleration can be provided by studying, for given initial and final configurations of a system, a $k$ -dependence of a probability distribution of time interval (number of steps in the discrete time dynamics) characterising trajectories connecting these configurations. In the literature one usually considers the first moment of this probability distribution, the mean time interval. For Markovian dynamics a general way of computing it is described in the literature on Markov chains, see e.g. \cite{Kemeny1976,Grinstead1997}. Another popular strategy of addressing this issue is to use the Langevin  \cite{bouchaud2013crises}  or Fokker-Planck equation formalism \cite{Risken1996}, the latter employed, in particular, in the analysis of metastable state decay rate in the self-excited Ising game in  \cite{antonov2021self,antonov2023transition}.

The plan of the paper is the following.

In Section 1 a $k$-flip Ising game is described. In paragraph 1.1 a general description of the $k$-flip dynamics. In paragraph 1.2 we describe an analytical calculation of the transition matrix of the considered game for arbitrary noise. In paragraph 1.3 we expressions for equilibrium distribution and the corresponding effective potential are given. In paragraph 1.4 a dynamical evolution in terms of a random walk in $\varphi=N^+/N$ is described and explicit expressions for $\langle \Delta \varphi \rangle$ and $\langle (\Delta \varphi)^2 \rangle$
are presented. In paragraph 1.5 a thermodynamic limit $N \to \infty$, $\rho = k/N={\rm const}$ is analysed. In paragraph 1.6 explicit expressions for mean transition time and mean squared transition time for arbitrary trajectories are calculated.

In Section 2 an analysis of properties of first hitting time for sample trajectories is given. In paragraph 2.1 general remarks on the considered problem are made. In paragraph 2.2 calculation of first two moments of first hitting time distribution in the framework of Markov chain formalism is discussed. In paragraph 2.3 we consider the solutions of the stationary distribution in Markov chain formalism. In paragraph 2.4 we describe the choice of the sample trajectories. In paragraph 2.5 parametric dependence of the first two moments of the first hitting time distribution for sample trajectories is discussed. An existence of the minimum in their dependence on $k$ is established, its origin argued to be in the competition of $k$-dependent confining and diffusional influences. In paragraph 2.6 numerical simulations supporting the results obtained are presented. 

Conclusions are presented in the corresponding section.

Appendices A and B provide details on the analytical calculation of $\langle \Delta \varphi \rangle$ and $\langle \Delta \varphi^2 \rangle$ correspondingly. Appendix C gives a detailed exposition of the calculation of the second moment of the first hitting time distribution.

\section{\label{sec:level2}A $k$ - flip dynamics}

\subsection{General description}

Let us consider a Markovian probabilistic dynamical evolution of a system characterised by a set of $N$ interacting binary variables ${\bf s} = (s_1, \dots,s_N)$, $\{ s_i = \pm 1\}$ in discrete time
\begin{equation}
{\bf s}(t) \overset{\Omega}{\longrightarrow} {\bf s} (t+1) \overset{\Omega}{\longrightarrow} {\bf s} (t+2) \overset{\Omega}{\longrightarrow} \dots 
\end{equation}
where, in a general case, $\Omega$ is a $2^N \times 2^N$ transition matrix such that
\begin{equation}
\Omega_{{\bf s}',{\bf s}''} = {\rm Prob} ({\bf s}' \longrightarrow {\bf s}'')
\end{equation}
A transition matrix $\Omega$ fully characterises the $k$-flip dynamics in which
\begin{itemize}
\item at each time step $t$ one randomly samples $k$ variables $i_1, \dots,i_k$  
\item for each $i \in (i_1, \dots,i_k)$ the variable $s_i$ flips, $s_i \to -s_i$, with a probability $p_i (s_i \to -s_i \vert {\bf s}_{-i} (t-1))$ that depends on the configuration of other variables at the previous time step ${\bf s}_{-i} (t-1)$. 
\end{itemize}

The interaction driving this evolution is assumed to be such that at any time step $t$ for any $i$ the interaction of $s_i$ with the rest of a system is characterised by the quantity $H^{\rm tot}_i s_i $ where 
\begin{equation}
H^{\rm tot}_i = \sum_{j \neq i} J_{ij} s_j (t-1) + H_i 
\end{equation}  
where $\{ J_{ij} \}$ are matrix elements of the interaction coupling matrix and $\{ H_i \}$ parametrise local idiosyncratic influences. We have therefore
\begin{equation}
p_i (s_i \to -s_i \vert {\bf s}_{-i} (t-1)) = p_i (s_i \to -s_i \vert (\sum_{j \neq i} J_{ij} s_j (t-1) + H_i) s_i  )
\end{equation}
The concrete functional form of flip probabilities depends on particular assumption on the structure of noise. For  given coupling matrix $J$ and assumptions on noise leading to the same expressions for flip probabilities $\{ p_i (-s_i \to s_i \vert {\bf s}_{-i}) \}$ the above-described description is valid for all the three Ising-type problems mentioned in the Introduction, i.e. to the Ising game, (generalised) Ising model and Little or Hopfield model. 

For the Ising game the flip $-s_i \to s_i$ is fully determined by the corresponding utility
\begin{equation}
U_i (s_i) = H^{\rm tot}_i  (t-1) s_i +  \varepsilon_{s_i} \equiv \left( \sum_{j \neq i} J_{ij} s_j (t-1) + H_i \right) s_i + \varepsilon_{s_i} 
\end{equation}
where $\varepsilon_{s_i} $ is a strategy - dependent random variable sampled from a noise distribution $f_i (\varepsilon_{s_i} )$. A probability of choosing $s_i$ is then 
\begin{equation}
p_i(s_i \vert H^{\rm tot}_i ) = {\rm Prob} [U_i (s_i) > U_i (-s_i) ] =  {\rm Prob} [\varepsilon_{-s_i} - \varepsilon_{s_i} < 2  H^{\rm tot}_i s_i]
\end{equation}
For the Gumbel noise
\begin{equation}\label{Gumbel}
f_i(\varepsilon_s) = \beta_i e^{-\beta_i \epsilon_s + e^{- \beta_i \epsilon_s}},
\end{equation}
where $\beta_i = 1/T_i$ parametrises the noise strengths, we get
\begin{equation}
p_i(s_i \vert H^{\rm tot}_i ) = \frac{e^{\beta_i H^{\rm tot}_i s_i}}{2 \cosh [ \beta_i H^{\rm tot}_i]}
\end{equation}

In what follows we shall restrict our consideration to the simplest case of a complete graph topology and, $\forall i,j \in 1, \dots, N$, uniform interactions with $J_{ij}=J>0$ and  homogeneous external influence and noise with $H_i = H$ and  $f_i (\varepsilon_s) = f(\varepsilon_s \vert \beta)$. In this case at large $N$ the utility of choosing $s$ is the same for all nodes
\begin{equation}
U_i (s_i) =  \left[ \frac{1}{N} \sum_{j \neq i} s_j + H \right] s_i + \varepsilon_{s_i} \underset{N \gg 1}{\simeq} \left[ \frac{1}{N} \sum_{j} s_j + H \right] s_i+ \varepsilon_{s_i}
\end{equation}
and so is thus that the choice probability
\begin{equation}
p_i(s) = p(s \vert m ), \;\;\; m = \frac{1}{N} \sum_{j} s_j \;\;\; \forall i \in 1, \dots, N
\end{equation} 
The choice probability does in this case depend only on the average choice
\begin{equation}
m =  \frac{1}{N} \sum_{j} s_j = \frac{N^+ - N^-}{N} = \frac{2 N^+ - N}{N} \equiv 2 \varphi -1
\end{equation}
where $N^\pm$ denote the number of nodes with $s=\pm 1$ respectively and we have defined 
\begin{equation}\label{defvarphi}
\varphi = N^+/N
\end{equation}
 Let us also define
\begin{equation}\label{ppgen}
p_+ (\varphi) \equiv p(s = 1\vert 2 \varphi - 1 )
\end{equation}

We see therefore that a state of a system ${\bf s}(t)$ is fully characterised by $N^+ (t)$ or, equivalently, $\varphi(t)$ and its evolution by specifying $p_+ (\varphi)$ so that 
\begin{equation}
N^+(t) \overset{\Omega^{(k)}}{\longrightarrow} N^+ (t+1) \overset{\Omega^{(k)}}{\longrightarrow} N^+ (t+2) \overset{\Omega^{(k)}}{\longrightarrow} \dots 
\end{equation}
where $\Omega^{(k)}$ is a $(N+1) \times (N+1)$ transition matrix for a $k$-flip game. 

\subsection{Transition matrix}

In this paragraph we calculate the exact analytical expression for the transition matrix $\Omega^{(k)}$.  

At any time step decisions of $k$ chosen agents result in one-step transitions of the form
\begin{equation}\label{transnp}
N^+ \;  \rightarrow \; N^+ + n, \;\;\; n  \in \left[- \min(k,N-N^+),\min(k,N^+) \right]
\end{equation}
taking place with probabilities $\omega^{(k)} (N^+ \; \rightarrow \; N^+ + n)$. In what follows we shall use the indices $i,j=0,\dots,N$ for indexing the $N+1$ generically possible initial and final configurations of such one-step transitions  with $N^+ = 0, \dots ,N$. The one-step evolution  is thus fully described by the transition matrix $\Omega^{(k)}$ with matrix elements
\begin{equation}
\Omega^{(k)}_{ij} = \omega^{(k)} (N^+ = i \; \rightarrow \; N^+ + n = j).
\end{equation}
Due to the myopic nature of agent's decisions the resulting dynamics of the $k$-flip Ising game is a Markovian one-dimensional random walk composed by one-step transitions \eqref{transnp} taking place with probabilities $\omega^{(k)} (N^+ \; \rightarrow \; N^+ + n)$.

The possible transitions for one agent are of the four following types: $(-1 \rightarrow +1), (-1 \rightarrow -1), (+1 \rightarrow +1), (+1 \rightarrow -1)$. Let us denote by $x_1,x_2,x_3,x_4$, $x_1+x_2+x_3+x_4=k$ the numbers of such events in a composition of the choices of $k$ agents at a given time step. The corresponding probability is then 
\begin{equation}
\frac{N^{-}!}{(N^{-} - x_1 - x_2)!}\frac{N^{+}!}{(N^{+} - x_3 - x_4)!}\frac{(N-k)!}{N!}.
\end{equation}
For a transition \eqref{transnp} one can express $n$ as $n=x_1-x_4$, so that

\begin{align}
& \omega^{(k)} (N^{+} \rightarrow N^+ +n) =  \sum_{\{x\}} \frac{k!}{x_1!\ x_2!\ x_3!\ x_4!} p_+^{x_1 + x_3} (1-p_+)^{x_2 + x_4} \nonumber \\
&\times \frac{N^{-}!}{(N^{-} - x_1 - x_2)!} \ \frac{N^{+}!}{(N^{+} - x_3 - x_4)!}\frac{(N-k)!}{N!} \nonumber \\
&\times {\mathbb I}[n,x_1-x_4] {\mathbb I}[k,x_1+x_2+x_3+ x_4] {\mathbb I}[s,x_2+ x_4], 
 \label{omegak1}
\end{align}

where the sum is taken over all possible $x_1, x_2, x_3, x_4 \leqslant k$ satisfying the conditions $x_1+x_2+x_3+x_4=k$ and $n=x_1-x_4$  and we have also introduced a new variable $s=x_2+x_4$.
Summing over $x_1,x_3,x_4$ in \eqref{omegak1} we get

\begin{align}
\omega^{(k)} (N^{+} \rightarrow N^{+} + n) &= \sum_{s, x_4} \frac{k!}{(x_4 + n)!\ (s-x_4)!\ (k - s - x_4 - n)!\ x_4!} \ p_+^{k - s} (1-p_+)^{s}  \nonumber \\
&\times \frac{N^{-}!}{(N^{-} - n - s)!}\frac{N^{+}!}{(N^{+} - k + s + n)!}\frac{(N-k)!}{N!}.
 \label{omegak2}
\end{align}

Summing over $x_4$ in \eqref{omegak2}\footnote{This can be done by temporarily shifting $n$ to the domain $n = 0, 1, ... , 2k$ by replacing $n \rightarrow n-k$ and expressing the summand in \eqref{omegak2} through binomial coefficients} we get
\begin{align}
\omega (N^{+} \rightarrow N^{+} + n) &= \sum_{s = \max(0, -n)}^{\min(k, k - n)} \binom{N - k}{N^{-} - s - n} 
\binom{k}{s} \binom{k}{n + s} \nonumber \\
&\times p_+^{k - s} (1-p_+)^{s}  \frac{N^{+}!N^{-}!}{N!}.
\end{align}

It is convenient to rewrite the lower and upper limits of the sum by introducing  the variables
\begin{equation}
\Delta^{\pm} = \frac{1}{2} (|n| \pm n),
\end{equation}
Then 
\begin{align}\label{omegak3}
\omega^{(k)} (N^{+} \rightarrow N^{+} + n) &= \sum_{s = 0}^{k - |n|} \binom{N - k}{N^{-} -s - \Delta^{+}} \binom{k}{s} \binom{k}{s + |n|} \nonumber \\
&\times  p_+^{k - s - \Delta^{-}}(1-p_+)^{s + \Delta^{-}} \frac{N^{+}!N^{-}!}{N!}.
\end{align}

The final expression for the matrix elements of transition matrix $\Omega^{(k)}_{ij}$ thus reads
\begin{align}\label{omegak}
\Omega^{(k)}_{ij} \equiv \omega^{(k)} (i \rightarrow j) &= \sum_{s = 0}^{k - |j - i|} \binom{N - k}{N - i - s - \Delta^{+}} \binom{k}{s} \binom{k}{s + |j - i|} \nonumber \\
&\times p_+^{k - s - \Delta^{-}}   (1-p_+)^{s + \Delta^{-}} \frac{i! (N - i)!}{N!},
\end{align}
where $p_+ \equiv p_+(\varphi = i / N)$.

Let us now discuss the two limiting cases of $k=1$ and $k=N$:
\begin{itemize}
\item In the 1-flip game with $k=1$ the only possible transitions are $N^+ \rightarrow N^+ \pm 1$ taking place with the following rates, see e.g.\cite{blume2003equilibrium,bouchaud2013crises,Leonidov2024}:
\begin{align}
\omega^{(1)} (N^+ \rightarrow N^+ + 1) = (1-\varphi) p_+ (\varphi), \;\;\; \omega^{(1)} (N^+ \rightarrow N^+ - 1) = \varphi (1 -  p_+ (\varphi) ),
\end{align}
\item In the k-flip game with $k=N$ the expression \eqref{omegak} simplifies to 
\begin{equation}
\Omega^{(N)}_{ij} = \binom{N}{N - i - \Delta^+} \binom{N}{N - i + \Delta^-}  p_+^{j} (1-p_+)^{N - j} \frac{i! (N - i)!}{N!}.
\end{equation}
It can be seen that, regardless of the sign of $j - i$, we get the same expression
\begin{align}
\Omega^{(N)}_{ij} &= \binom{N}{j}  p_+^{j} (1-p_+)^{N - j}.
\end{align}
Therefore, for  $k = N$, the expression for the transition probability corresponds to a binomial distribution.
\end{itemize}

\subsection{Equilibrium distribution}

The availability of the exact expression \eqref{omegak} for the the transition matrix $\Omega^{(k)}$ allows to find the asymptotic equilibrium distribution $\pi^{(k)}_{\varphi}$ for the $k$-flip dynamics under consideration by solving the equation
\begin{equation}\label{eqdistk}
\pi^{(k)}_{\varphi} \Omega^{(k)} = \pi^{(k)}_{\varphi}.
\end{equation}
For the subsequent analysis It turns out convenient to parametrise the distribution $\pi^{(k)}_{\varphi}$ by introducing the effective potential $V^{(k)}_{\rm eff} (\varphi)$ through
\begin{equation}\label{veffk}
V^{(k)}_{\rm eff} (\varphi) = - \ln \pi^{(k)}_{\varphi}
\end{equation}

\subsection{Random walk in $\varphi$}

In the previous paragraph we have derived the exact expression \eqref{omegak} for the matrix elements of the transition matrix $\Omega^{(k)}$ which can be used for studying important characteristics of the corresponding Markovian random walk in $N^+$ or, equivalently, in $\varphi=N^+/N$ composed by one-step transitions $\varphi \to \Delta \varphi$ taking place with probabilities $\omega^{(k)} (\varphi \; \rightarrow \; \varphi + \Delta \varphi)$
\begin{equation}
\varphi \; \rightarrow \; \varphi + \Delta \varphi, \;\;\;  \Delta \varphi \in \left[ \min ( \rho,1-\varphi ), \min ( \rho,\varphi ) \right], \;\;\; \rho \equiv \frac{k}{N},
\end{equation}

The distribution of step-length $n$ (or $\Delta \varphi = n/N$) is completely determined by the corresponding matrix elements of transition matrix $\Omega^{(k)}_{ij} = \omega^{(k)} (\varphi = i/N \; \rightarrow \; \varphi' = j/N)$ \eqref{omegak}. Of crucial importance for the subsequent analysis of first hitting time for trajectories of interest $i \rightarrow j$ are the expressions for the first two moments of the probabilistic distribution of $\Delta \varphi$,  $\langle \Delta \varphi \rangle$ and $\langle (\Delta \varphi)^2 \rangle$. The initial expression for first moment $\langle \Delta \varphi \rangle$ reads
\begin{align}
\langle\Delta \varphi \rangle &= \sum_{\{x\}} \frac{k!}{x_1!\ x_2!\ x_3!\ x_4!} \ p_+^{x_1 + x_3} (1-p_+)^{x_2 + x_4} \nonumber \\
&\times \frac{N^{-}!}{(N^{-} - x_1 - x_2)!}  \frac{N^{+}!}{(N^{+} - x_3 - x_4)!} \frac{(N-k)!}{N!} \ \frac{1}{N} (x_1 - x_4),
\label{first_monent}
\end{align}
where all notations correspond to those in (\ref{omegak1}). After performing the summation in \eqref{first_monent} (for details see Appendix A), we obtain
\begin{equation}
\langle\Delta \varphi \rangle = \frac{k}{N} (p_+(\varphi) - \varphi).
\label{mean_dphi}
\end{equation}
Thus, the average bias grows linearly with $k$. 

For the second moment we have  
\begin{align}
\langle (\Delta \varphi)^2 \rangle &= \sum_{\{x\}} \frac{k!}{x_1!\ x_2!\ x_3!\ x_4!} \ p_+^{x_1 + x_3} (1-p_+)^{x_2 + x_4} \nonumber \\
&\times \frac{N^{-}!}{(N^{-} - x_1 - x_2)!} \frac{N^{+}!}{(N^{+} - x_3 - x_4)!}\   \frac{(N-k)!}{N!} \ \frac{1}{N^2} (x_1 - x_4)^2.
\label{second_monent}
\end{align}
This sum is calculated using methods similar to those used for calculating (\ref{mean_dphi}) (for detailed calculations see Appendix B). We get the following expression:
\begin{equation}
\langle (\Delta \varphi)^2 \rangle  = \frac{k}{N^2} \Bigg( p_+^2 (k-1) + p_+ (1 - 2k\varphi) + \varphi\frac{-k + N + (k - 1)N\varphi}{N-1} \Bigg).
\label{mean_dphi2}
\end{equation}
Of interest here is a non-trivial dependence of $\langle (\Delta \varphi)^2 \rangle$ on $k, N$, which in the general case does not reduce to a dependence  on the ratio $\rho = k/N$.

Using the formulae (\ref{mean_dphi}) and (\ref{mean_dphi2}) one can also obtain an explicit expression for the standard deviation of $\Delta \varphi$:
\begin{equation}
\sigma_{\Delta \varphi} = \sqrt{\langle (\Delta \varphi)^2 \rangle - \langle \Delta \varphi \rangle^2}.
\label{sigma_dphi}
\end{equation}

\subsection{Thermodynamic limit $N \rightarrow \infty$, $\rho=k/N={\rm const}$}

In the limit $N \rightarrow \infty$, $\rho={\rm const}$  the variance (\ref{sigma_dphi}) is found to exhibit the following asymptotic behavior:
\begin{equation}
\sigma_{\Delta \varphi}^2 = \frac{\rho}{N} \big[ p_+ (1 - p_+) + \varphi(1-\varphi)(1-\rho) \big] + O (1/N^2).
\label{sigma_approx}
\end{equation}
Thus, the variance is of order $1/N$ and the mean displacement is given by $\langle (\Delta \varphi) \rangle = \rho (p_+ - \varphi)$. Consequently, the distribution of the step $\delta \varphi = (j - i) / N$ is concentrated about $\varphi' = \varphi + \langle (\Delta \varphi) \rangle = \varphi +  \rho (p_+ - \varphi)$, with a narrow spread characterized by the variance (\ref{sigma_approx}).

Let us introduce the transition probability density $f(\varphi' | \varphi)$ from state $\varphi$ to state $\varphi$, satisfying the normalization condition
\begin{equation}
\int_0^1 d\varphi' f(\varphi' | \varphi)  = 1.
\end{equation}
Then, for large $N$, the probability density takes the approximate form
\begin{equation}
f(\varphi' | \varphi) \approx \frac{1}{\sqrt{2 \pi}\sigma_{\Delta \varphi}} \exp \Bigg( - \frac{1}{2} \frac{(\varphi' - \varphi - \rho(p_+ - \varphi))^2}{\sigma_{\Delta \varphi}^2} \Bigg).
\label{omega_approx}
\end{equation}
In the limit $N \rightarrow \infty$ we get
\begin{equation}
\lim_{N \rightarrow \infty} f(\varphi' | \varphi) = \delta(\varphi' - \varphi - \rho(p_+ - \varphi)).
\end{equation}
Hence, for large $N$, the system dynamics approach a deterministic regime characterized by jumps $\varphi' - \varphi = \rho(p_+ - \varphi)$, accompanied by weak diffusion in the vicinity of the state  $\varphi'$.

It is important to note that the approximation (\ref{omega_approx}) accurately reproduces the center of the distribution; however, it significantly transforms the tails, rendering them considerably heavier compared to the binomial nature of the original distribution (\ref{omegak}). This discrepancy may substantially affect the system dynamics, particularly the rate of convergence to the stationary state. Consequently, the mean transition times between system states may deviate markedly from those obtained using the exact distribution (\ref{omegak}).

\subsection{Transition times in Markov chain formalism}

One of the main questions studied in this work is a consideration of $k$-dependence of first hitting times characterising some particular trajectories.  The first hitting time characterising a trajectory $i \rightarrow j$ is defined as a number of steps $\tau_{ij}$ required for the system to reach the state $j$ starting from the state $i$ in the $k$-flip game under consideration
\begin{equation}
\tau_{ij} : \;\; i \overset{\omega_{ij}^{(k)}}{\longrightarrow} \;\; j \;\; \leftrightarrow \;\; 
\varphi_i \overset{\omega_{ij}^{(k)}}{\longrightarrow} \;\; \varphi_j 
\end{equation}

For trajectories $i \rightarrow j$ generated by the random walk evolution of a $k$ - flip Ising game the variable $\tau_{ij}$ is stochastic and thus characterised by some probability distribution ${\cal P}^{(k)} (\tau_{ij})$. For a given transition matrix $\Omega^{(k)}_{ij}$ its moments can be directly assessed using the Markov chain theory description of the random walk, see e.g. \cite{Kemeny1976,Grinstead1997}. In particular, the first two moments satisfy the following recurrent equations:
\begin{eqnarray}
\langle \tau_{ij} \rangle & = & 1 \cdot \Omega^{(k)}_{ij} + \sum_{n \neq j} \Omega^{(k)}_{in}  \langle (\tau_{nj} + 1) \rangle  =  1 + \sum_{n \neq j} \Omega^{(k)}_{in}  \langle \tau_{nj} \rangle, \label{mtau1} \\
\langle (\tau_{ij})^2 \rangle &=& 1 \cdot \Omega^{(k)}_{ij} + \sum_{n \neq j} \Omega^{(k)}_{in}  \langle (\tau_{nj} + 1)^2 \rangle \nonumber \\
& = & 1 + \sum_{n \neq j} \Omega^{(k)}_{in}  \langle (\tau_{nj})^2 \rangle + 2 \sum_{n \neq j} \Omega^{(k)}_{in}  \langle \tau_{nj} \rangle. \label{mtau2}
\end{eqnarray}

Equations (\ref{mtau1},\ref{mtau2}) can be compactly written in terms of the fundamental matrix $Z^{(k)}$
\begin{equation}\label{defZ}
Z^{(k)} = I + \sum_{n=1}^\infty (\Omega^{(k)} - \Pi^{(k)})^n = (I - \Omega^{(k)} + \Pi^{(k)})^{-1}.
\end{equation}
where $\Pi^{(k)}$ is a matrix in which all the rows are equilibrium distribution $\pi^{(k)}$ satisfying the global balance equation \eqref{eqdistk} and $I$ is a unit matrix. Let us define the matrices
\begin{equation}
T: \;\; T_{ij} = \langle \tau_{ij} \rangle, \;\;\; T^{(2)}: \;\; T_{ij} = \langle (\tau_{ij})^2 \rangle
\end{equation}
Then
\begin{eqnarray}
T & = & (E\diag Z^{(k)} - Z^{(k)} + I) D^{(k)}, \label{m_ij} \\
T^{(2)} &=& 2 T D^{(k)} \diag Z^{(k)}- 3T + 2 [ I + E\diag((Z^{(k)})^2) - (Z^{(k)})^2 ]D^{(k)}, \label{var_tau_main}
\end{eqnarray}
where $D^{(k)} = {\rm diag} (1/\pi^{(k)})$ and $E$ is a matrix with all its matrix elements equal to one. The equation \eqref{m_ij} is the known equation for the matrix of mean first hitting times in \cite{Kemeny1976}. The detailed derivation of \eqref{var_tau_main} is presented in the Appendix C.

\section{Decay of metastable states in the  $k$ - flip Ising dynamics}

\subsection{General setting}

\begin{figure}[ht]
	\center{\includegraphics[width=1\linewidth]{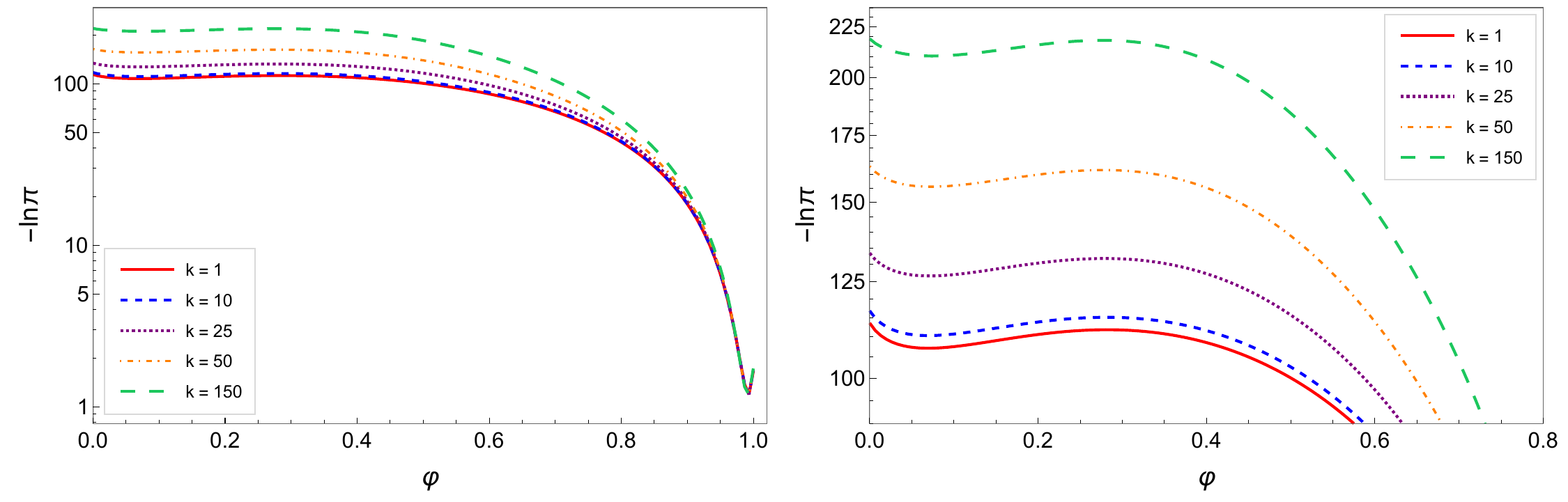}}
	\caption{\small Plots of the potential $V_{\varphi} (k, N, \beta, H) = - \ln \pi^{(k)}_\varphi (k, N, \beta, H)$ as a function of state $\varphi$ for $\beta = 1.9, J = 1, H = 0.8H^{*}, N = 150$ and $k = 1$ (solid line), $k = 10$ (dashed line), $k = 50$ (dotted line), $k = 100$ (dash-dotted line), $k = 150$ (long-dashed line). The right plots correspond to an enlarged scale displaying the metastable state.}
	\label{pi_dist}
\end{figure}

All the results of the previous section are valid for arbitrary noise distribution $f(\varepsilon_s)$.  A convenient context for discussing the main problem addressed in this section, the analysis of decay of metastable states in the $k$-flip dynamics on complete graphs, is ensured by considering the Gumbel noise \eqref{Gumbel} so that at any given time $t$
\begin{equation}
p(s \vert t) = \frac{e^{\beta [H + J (2 \varphi(t) - 1)]s }}{e^{\beta [H + J (2 \varphi(t) - 1)]s} + e^{-\beta [H + J (2 \varphi(t) - 1)]s}}
\end{equation}
In particular, for  $p_+ (\varphi (t)) \equiv p(s = 1 \vert t)$ we get
\begin{align}\label{pp}
p_+ (\varphi (t)) = \frac{e^{\beta [H + J (2 \varphi(t) - 1)] }}{2 \cosh [\beta H + \beta J (2 \varphi(t) - 1)]}.
\end{align}
Plugging expression \eqref{pp} into a general formula for the transition matrix \eqref{omegak}, one gets a particular expression for the transition matrix $\Omega^{(k)} (\beta,H)$ and, through solving \eqref{eqdistk}, one determines the equilibrium distribution $\pi^{(k)}_{\varphi} (k, N, \beta, H)$ and, therefore, the corresponding effective potential
\begin{equation}\label{veff_pi}
V^{(k)}_{\varphi} (N, \beta, J, H) = - \ln \pi^{(k)}_{\varphi} ( N, \beta, J, H).
\end{equation}
The shape of $V^{(k)}_{\varphi} (N, \beta, J, H)$ is illustrated, for several values of $k$, in Fig. ~\ref{pi_dist}, at fixed $\beta J$, $N$ and $H = 0.8H^{*}$ where $H^*$ is a critical external influence (field) corresponding to the end of the hysteresis region, see below (for details see \cite{Leonidov2024}).

\begin{figure}[H]
	\centering{\includegraphics[width=1\linewidth]{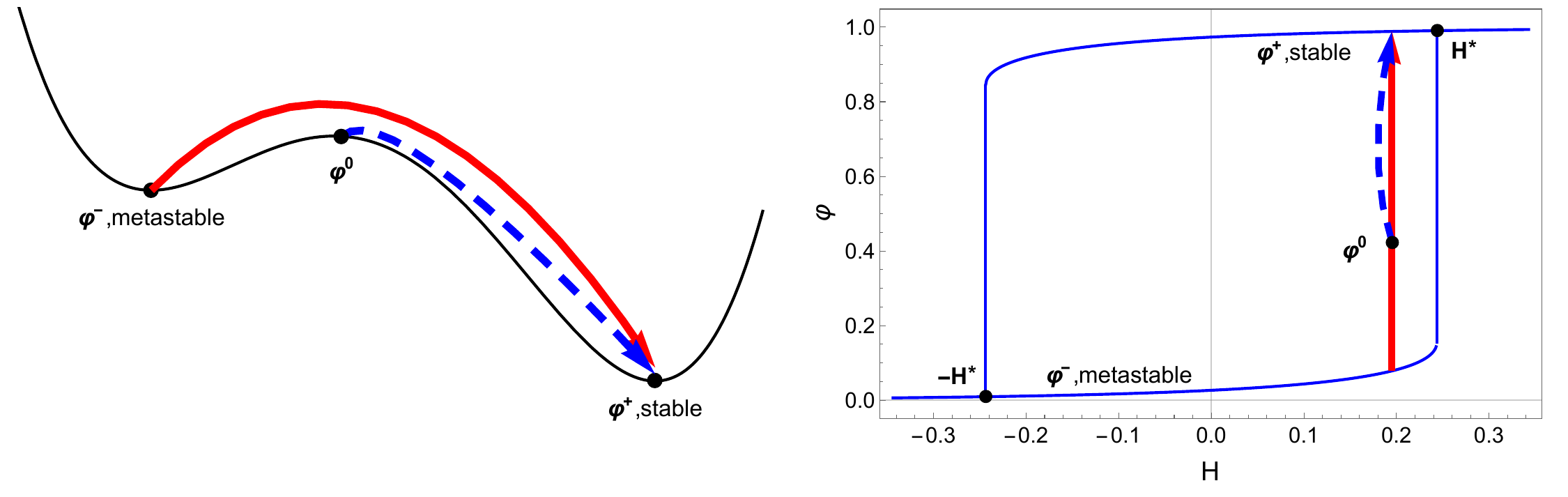}}
	\caption{\small The schematic structure of the effective potential $V_{\rm eff} (\varphi \vert \beta,H)$ (left) and the structure of solutions  (right) for the low-temperature phase. The red solid arrow shows the transition from the metastable state $\varphi^-(H)$ at the lower branch to the stable state $\varphi^+(H)$ at the upper branch. The blue dashed arrow shows the transition from unstable state $\varphi^0(H)$ to the stable state $\varphi^+(H)$. The parameters are  $\beta J = 1.9, \gamma=0.8$.}
	\label{pot_example}
\end{figure}

The effective potential $V^{(1)}_{\varphi}$ for the 1-flip dynamics describes the structure of conventional static equilibria in both the Ising model and Ising game on complete graphs, see e.g. \cite{blume2003equilibrium,bouchaud2013crises,Leonidov2024}. At low temperatures it is characterised by the coexistence of three equilibria $\varphi^{\pm} (H)$ and $\varphi^{0} (H)$ in the hysteresis region $H \in [-H^* (\beta J),H^*(\beta J)]$ (for details see \cite{Leonidov2024}). At $H>0$ the equilibrium $\varphi^+$ is stable and corresponds to the right deeper minimum of  $V^{(1)}_{\varphi}$ in Fig. ~\ref{pi_dist}, the equilibrium $\varphi^0$ is unstable and corresponds to its maximum and the equilibrium $\varphi^-$ is metastable and corresponds to its left shallow minimum. The decay of a metastable state $\varphi^-$ into the stable one $\varphi^+$ is schematically illustrated in 
Fig. ~\ref{pot_example} where we also show the corresponding trajectory on top of the hysteresis diagram.  

The $k$ - dependence of the decay rates for the trajectories $\varphi^- \to \varphi^+$ and $\varphi^0 \to \varphi^+$ is the main problem studied in this section. Let us stress that in general a transition from sequential $1$-flip game to a partially parallel $k$- flip one does significantly affect all its aspects including the very character of dynamical evolution. In the present paper we concentrate on one specific aspect of distinction between the two games, a difference in time required to complete a transition between two given states of a system. Naively one would expect that in the (partially) parallel game the evolution is faster and so is this transition. We will see that this naive expectation is not supported by the analysis presented below and that the picture is more nuanced.

\subsection{Decay times at sample trajectories}

Let us turn to an analysis of the properties of first passage times at the sample trajectories $\varphi^-,\varphi^0 \rightarrow \varphi^+$ described in the previous paragraph. In what follows we will use a notation $i^* \rightarrow j^*$ for both trajectories so that $i^*=\varphi^-, \varphi^0$ and $j^*=\varphi^+$.

\begin{figure}[H]
	\center{\includegraphics[width=0.6\linewidth]{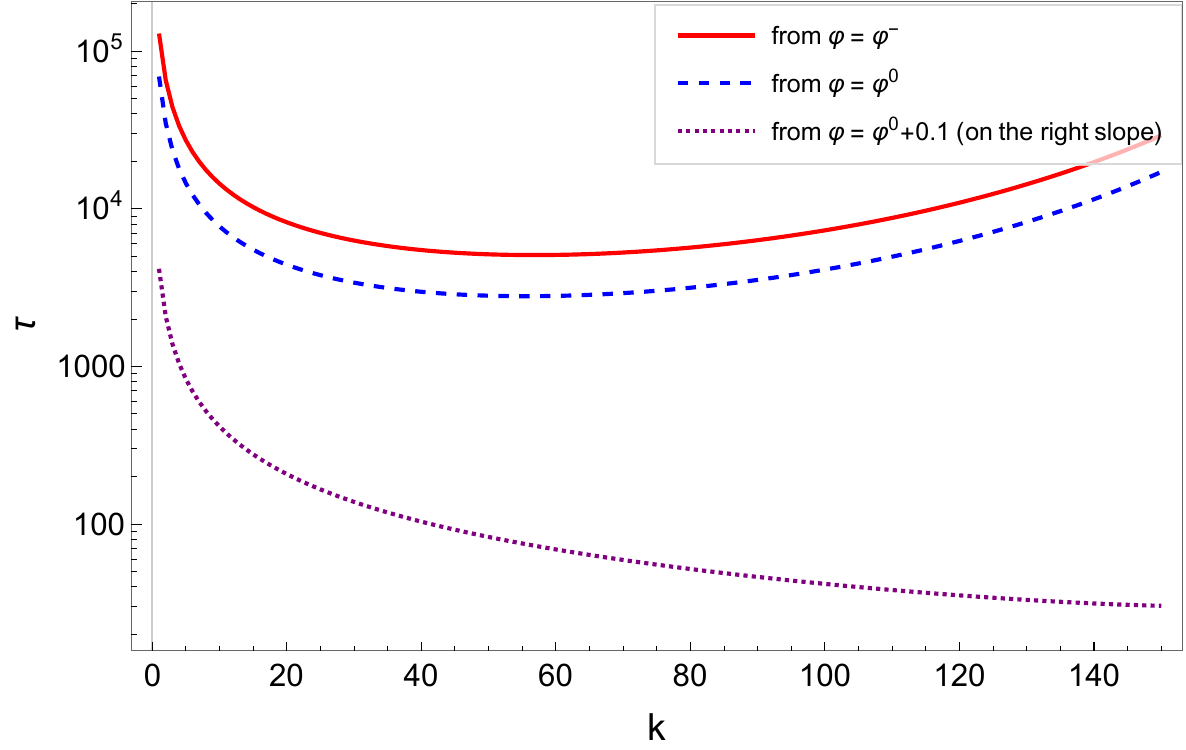}}
	\caption{\small Plots of mean first-hitting times as a function of $k$ for $\beta = 1.9, \gamma = 0.8$ and $N = 150$. The red solid (blue dashed) line corresponds to the red solid (blue dashed) arrow in Fig. \ref{pot_example}. The dotted line corresponds to the start point $\varphi = \varphi^0 + 0.1 (i^* = \varphi^0 + 0.1)$ on the right slop of the potential $V^{(k)}_{\varphi}$.}
	\label{tau_th_m0}
\end{figure}

The first quantity we'll analyse is the mean first passage time $T_{i^*j^*} = \langle \tau_{i^*j^*} \rangle$. We have from \eqref{m_ij}
\begin{equation}\label{T_main}
T_{i^*j^*}=\frac{Z_{j^*j^*}-Z_{i^*j^*}}{\pi_j^*}.
\end{equation}
Let us now consider, for the Gumbel noise \eqref{Gumbel}, the dependence of the mean first-hitting time $T_{i^* j^*}$  on the parameters $k, \beta, \gamma, N$. 

In Fig.~\ref{tau_th_m0} we show the plot the mean first-hitting times as a function of the number of simultaneously playing agents $k$ for the transitions $\varphi^- \rightarrow \varphi^+$ (red line) and $\varphi^0 \rightarrow \varphi^+$ (blue dashed line) for $N=150$.

One can see that for the trajectory from the point on the right slope of the potential $\varphi^0+ 0.1 \rightarrow \varphi^+$ the mean first hitting time $T_{i^*j^*}$ is a decaying function of $k$, while for the trajectory  $\varphi^- \rightarrow \varphi^+, \varphi^0 \rightarrow \varphi^+$ the plot $T_{i^*j^*} (k)$ shows an initial decay and then a growth after reaching a cetain critical value $k$.  For the transition $\varphi^- \rightarrow \varphi^+$ we denote this critical point as $k_{\rm min}$.

\begin{figure}[H]
	\center{\includegraphics[width=1\linewidth]{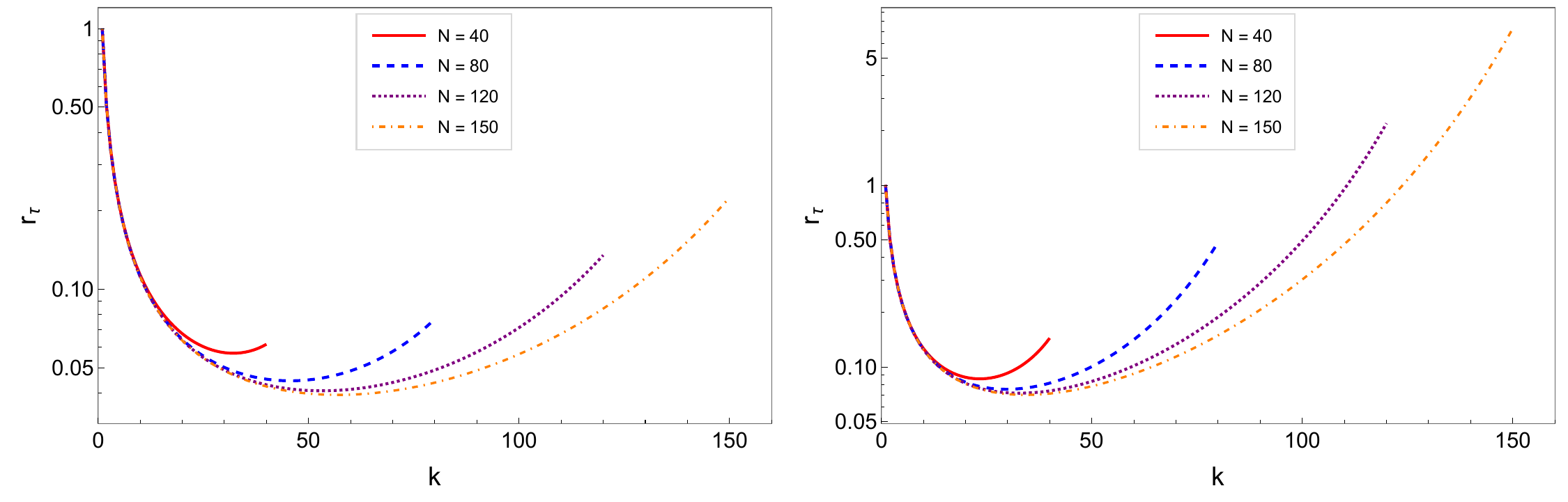}}
	\caption{\small Plots of \( r_\tau (\beta, \gamma, k, N)  \) as a function of $k$ for $N =40$ (solid line), $N =80$ (dashed line), $N = 120$ (dotted line), $N = 150$ (dash-dotted line) with \(\gamma = 0.8\) and $\beta = 1.9$ (left), $\beta = 2.5$ (right).}
	\label{tau_th_k}
\end{figure}

It is convenient to compare the results for the (partially) parallel $k$-flip Ising game at $k>1$ to those for the sequential one at $k=1$ by introducing the ratio
\begin{equation}
r_\tau (\beta, \gamma, k, N) = \frac{ T_{i^* j^*}(\beta, \gamma, k, N) }{T_{i^* j^*}(\beta, \gamma, k = 1, N)},
\end{equation}
The $k$-dependence of $r_\tau (\beta, \gamma, k, N) $ is shown, for the trajectory $\varphi^- \rightarrow \varphi^+$, for several values of $N$ and two values of $\beta$ in Fig.~\ref{tau_th_k}. The presence of a minimum at $k_{\rm min}$ is clearly seen. Notably, higher values of  $\beta$ amplify the growth rate of $r_\tau (\beta, \gamma, k, N)$ with $k$ at $k>k_{\rm min}$. A similar effect is observed for increasing $N$, though with an additional non-trivial feature. While one might intuitively expect the normalized transition rate $r_\tau(\beta, \gamma, \rho, N)$ to be independent of $N$ (i.e. depend only on $\rho$), the results reveal that $N$ plays a non-negligible role in the structure of the transition matrix $\Omega$.

\begin{figure}[H]
	\center{\includegraphics[width=1\linewidth]{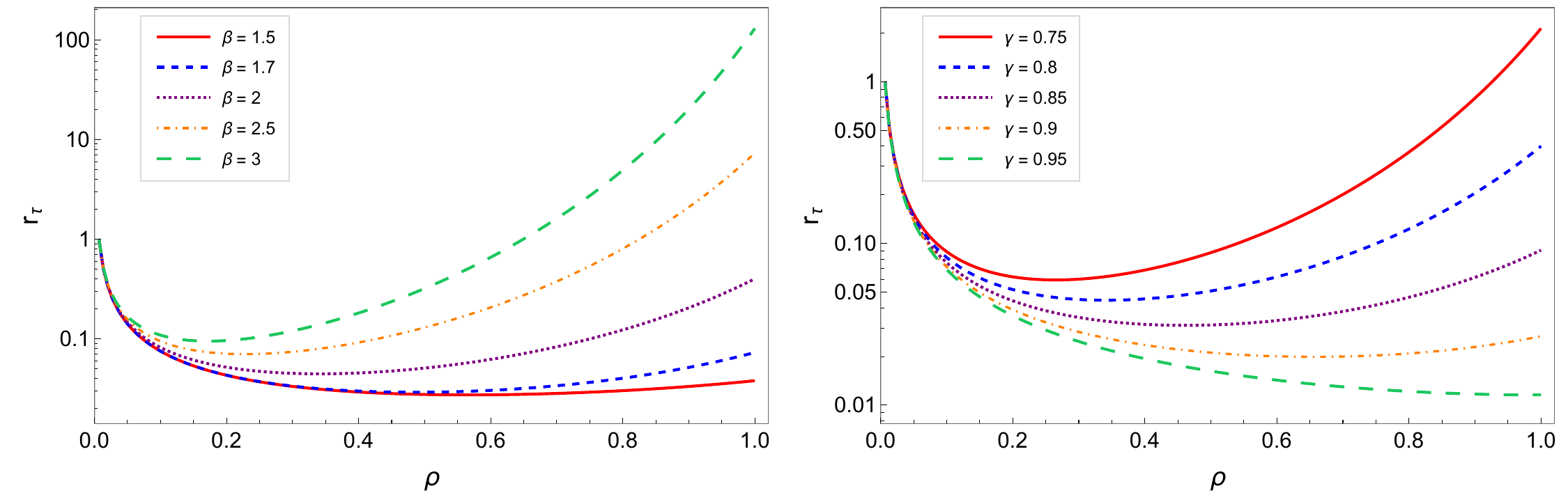}}
	\caption{\small Plots of \( r_\tau (\beta, \gamma, \rho, N)  \) as a function of $\rho$ with \(N = 150\) (1) for $\beta = 1.5$ (solid line), $\beta = 1.7$ (dashed line), $\beta = 2$ (dotted line), $\beta = 2.5$ (dash-dotted line), $\beta = 3$ (long-dashed line) and $\gamma = 0.8$ (left figure); (2) for $\gamma = 0.75$ (solid line), $\gamma = 0.8$ (dashed line), $\gamma = 0.85$ (dotted line), $\gamma = 0.9$ (dash-dotted line), $\gamma = 0.95$ (long-dashed line) and $\beta = 2$ (right figure).}
	\label{tau_th_rho}
\end{figure}

\begin{figure}
	\center{\includegraphics[width=1\linewidth]{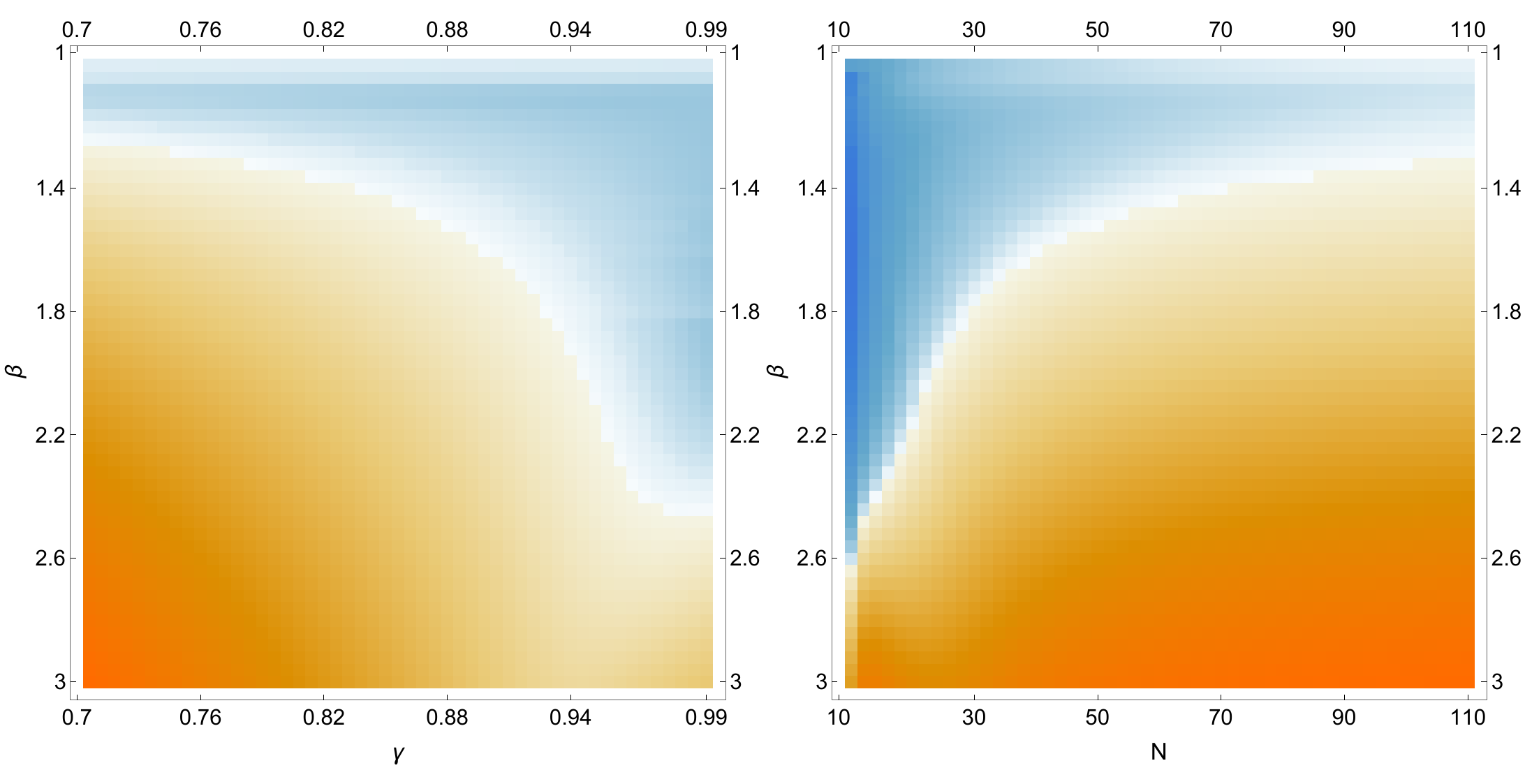}}
	\caption{\small Diagrams of \( \ln [r_\tau (\beta, \gamma, k = N, N) / r_\tau (\beta, \gamma, k = N - 1, N)]  \) in projections on $\beta \gamma$ ($N = 80$) (left) and $\beta N$ ($\gamma = 0.8$) (right).}
	\label{phase_diagram}
\end{figure}

\begin{figure}
	\center{\includegraphics[width=1\linewidth]{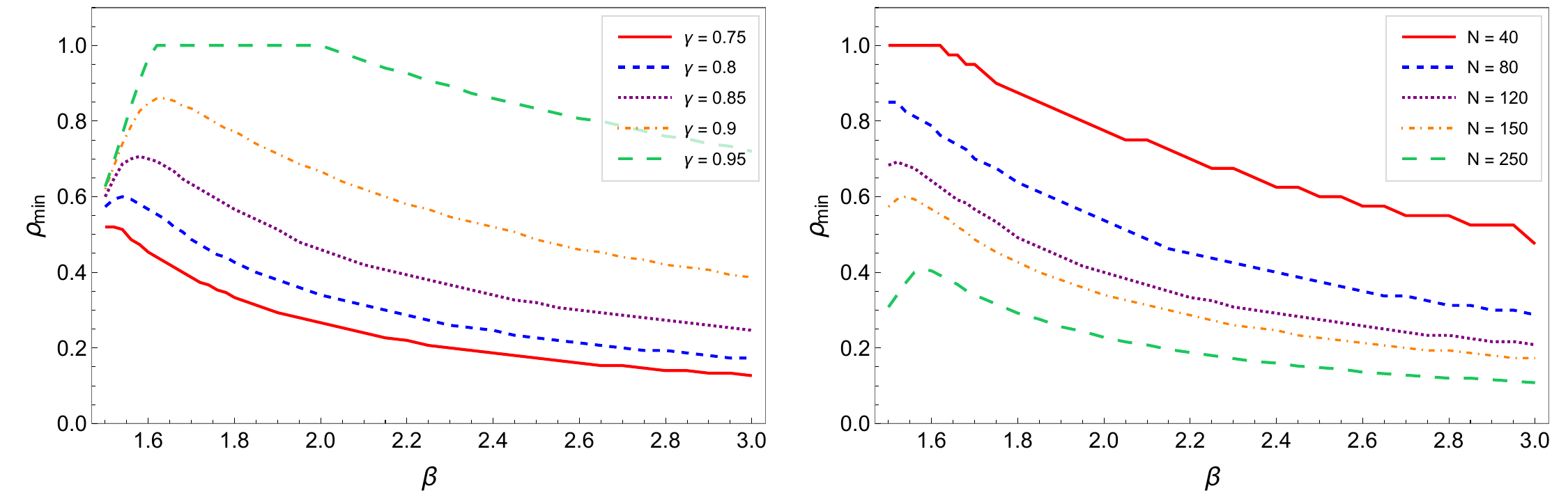}}
	\caption{\small Plots of the minimum position of mean first-hitting time $\rho_{\rm min}$ as a function of $\beta$ (1) for  $\gamma = 0.75$ (solid line), $\gamma = 0.8$ (dashed line), $\gamma = 0.85$ (dotted line), $\gamma = 0.9$ (dash-dotted line), $\gamma = 0.95$ (long-dashed line) and $N = 150$ (left figure); (2) for $N =40$ (solid line), $N =80$ (dashed line), $N = 120$ (dotted line), $N = 150$ (dash-dotted line), $N = 250$ (long-dashed line) and $\gamma = 0.8$ (right figure).}
	\label{rho_min}
\end{figure}

In Fig.~\ref{tau_th_rho} we plot the dependence of $r_\tau (\beta,\gamma, k, N)$ on $\rho=k/N$ for different $\beta$ and $\gamma$. As noted earlier, an increase in $\beta$ causes a growth of the mean transition time for $k > k_{\rm min}$. However, the dependence of the transition time for different $\gamma$ shows that for some $\beta, \gamma$ the minimum does not appear. That is, there is a regime with a minimum in the transition time, and a regime in which the transition time decreases monotonically with increasing $\rho$ (or $k$).

The phase diagram describing the coexistence of the two regimes is defined in the parameter space $N$, $\beta, \gamma$. Fig. \ref{phase_diagram} shows color maps projected onto $\beta\gamma$ and $\beta N$ planes, displaying the logarithmic ratio of the average transition times $\ln\Big[\langle\tau(k= N)\rangle / \langle\tau(k = N - 1)\rangle \Big]$. Areas with positive values of this ratio (corresponding to the existence of a minimum for $\langle\tau\rangle$) are highlighted in warm tones, while areas with negative values (characterizing the monotonous decrease of $\langle\tau\rangle$ at $\rho\rightarrow 1$) are highlighted in cold ones. The analysis of the phase diagram in the $\beta\gamma$ plane reveals the presence of a non-trivial transition between these two modes.

It is also of interest to study the dependence of the position of this minimum on the system parameters. In Fig.~\ref{rho_min}, we show the dependence of $\rho_{\rm min}=k_{\rm min}/N$ on the inverse temperature $\beta$. We see that with increasing $\beta$ the minimum shifts to lower values of $\rho$. At the same time, as can be seen from the plots \ref{tau_th_rho}, higher values of $\beta$ correspond to lower values of $r_\tau (\beta, \gamma, \rho_{\rm min}, N)$. The shift of $\rho_{\rm min}$ towards lower values is also due to a decrease in the parameter $\gamma$ and an increase in the size of the system $N$. It should also be noted that in Fig. \ref{rho_min} value of $\rho_{\rm min} = 1$ corresponds to a regime in which there is no minimum. In the limit $N \rightarrow \infty$, the asymptotic behavior of  $\rho_{\rm min}$ is such that $\rho_{\rm min} \sim 1/N$ ($\rho$ is strictly confined to the domain $\rho \geqslant 1/N$).

\begin{figure}[H]
	\center{\includegraphics[width=0.6\linewidth]{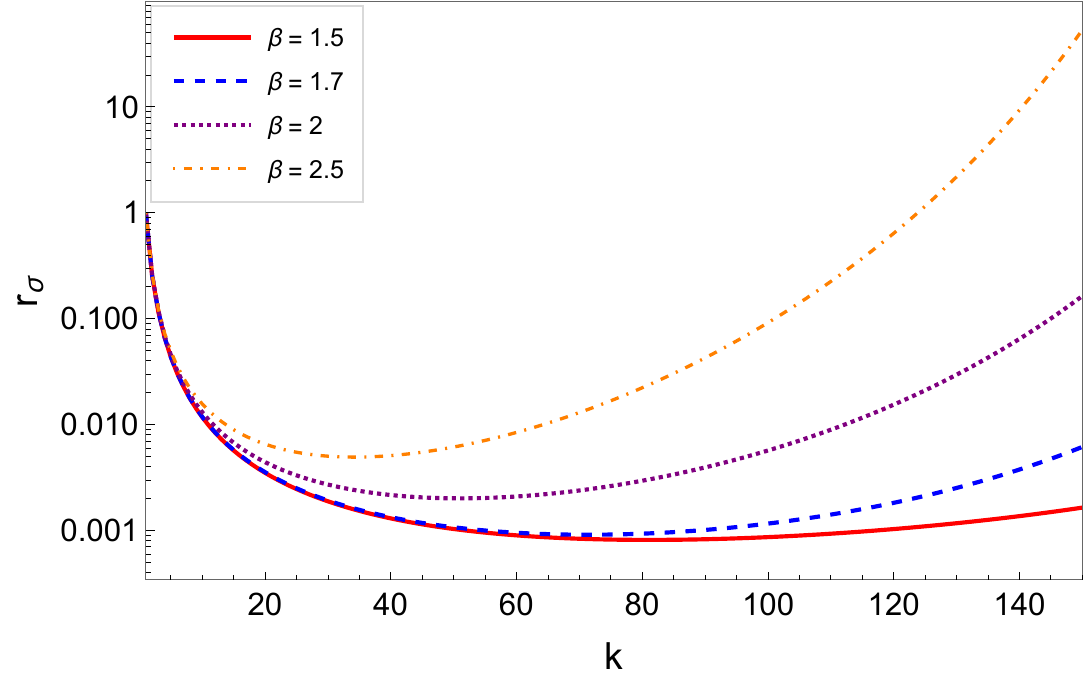}}
	\caption{\small  Plot of $r_\sigma (\beta, \gamma, k, N)$ as a function of $k$ with \(N = 150, \gamma = 0.8\) for $\beta = 1.5$ (solid line), $\beta = 1.7$ (dashed line), $\beta = 2$ (dotted line), $\beta = 2.5$ (dash-dotted line).}
	\label{var_tau_fig}
\end{figure}

It can be shown that higher-order moments of the first-hitting time distribution exhibit analogous behaviour. Let us focus on the variance of     the hitting time distribution 
\begin{equation}\label{var_tau_true}
{\rm Var}\; \tau_{ij} = T^{(2)}_{ij} - T_{ij}^2
\end{equation}
where the mean time $T_{ij}$ and mean squared time $T^{(2)}_{ij}$ can be calculated using the equations (\ref{m_ij},\ref{var_tau_main}). In Fig.~\ref{var_tau_fig} we plot the ratio
\begin{equation}
r_\sigma (\beta, \gamma, k, N) = \frac{ {\rm Var} \; \tau_{i^* j^*}(\beta, \gamma, k, N) }{{\rm Var} \; \tau_{i^* j^*}(\beta, \gamma, k = 1, N)}
\end{equation}
as a function of $\rho$ for several values of $\beta$ for the trajectory $\varphi^- \rightarrow \varphi^+$. The minimum of $r_\sigma (\beta, \gamma, k, N)$ at some $\rho_{\rm min}$ is clearly visible. 

\begin{figure}[ht]
	\center{\includegraphics[width=1\linewidth]{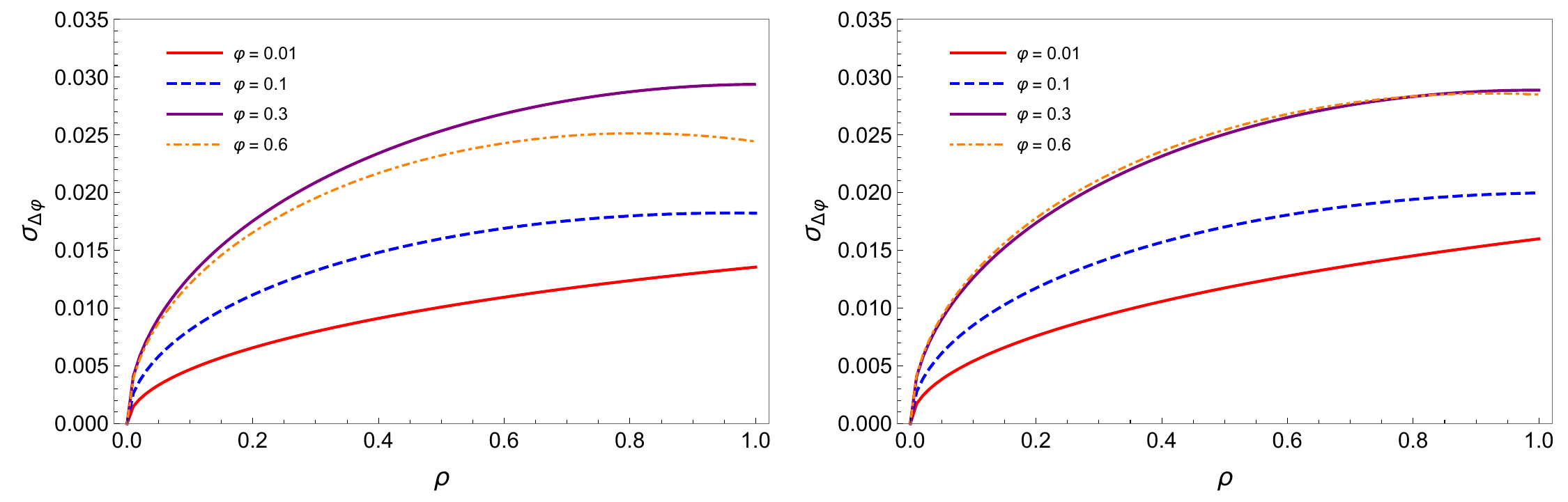}}
	\caption{\small Plot of $\sigma_{\Delta \varphi}$ as a function of $\rho = k/N$ for $\gamma = 0.8$ and $\beta = 1.9$ (left), $\beta = 1.5$ (right) with $\varphi = 0.01$ (solid line), $\varphi = 0,1$ (dashed line), $\varphi = 0.3$ (dotted line), $\varphi = 0.5$ (dash-dotted line).}
	\label{sigma_rho}
\end{figure}

To explain the presence of minimum of $r_\tau (\beta, \gamma, k, N)$ let us notice that any deviation from the stationary (or quasi-stationary) state induces a restoring force (\ref{mean_dphi}), which acts to shift the step-length distribution back towards the stationary (or quasi-stationary) state. The magnitude of this restoring force increases linearly with $k$. Considering the plots \ref{sigma_rho} of the standard deviation (\ref{sigma_dphi}) as a function of $k$, it is evident that its growth rate decelerates substantially with increasing $k$, i.e. its  growth rate is significantly higher for small values of $k$ as compared to larger ones. This leads to the conclusion that the emergence of a minimum in the transition time can be attributed to the competition between two opposing mechanisms. The first mechanism is the rapid increase in the width of the step-length distribution, corresponding to an enhancement of diffusion within the system. This factor acts as a mechanism that accelerates the escape from a metastable state. The second mechanism is associated with the mean displacement, or the restoring force. This mechanism, in contrast, provides an increasing confining effect as $k$ grows. For small $k$, the diffusion is the dominant factor, enabling the system to escape the metastable state more rapidly. However, at a specific point $k = k_{\rm min}$ (or $\rho = \rho_{\rm min}$), the influence of these two factors becomes comparable. Beyond this point, the confining effect of the restoring force surpasses the accelerating effect of diffusion.

In order to estimate $k_{\rm min}$, we can compare the growth rates of $\langle\Delta \varphi \rangle$ and $\sigma_{\Delta \varphi}$ with respect to $k$ at some intermediate point $\varphi = \varphi_{\rm mid}$ between the meta-stable state and the saddle point. This is achieved by taking the derivatives of both and setting them equal. For the derivative $\partial \langle \Delta \varphi \rangle/ \partial k$, we obtain
\begin{equation}
\frac{\partial \langle \Delta \varphi \rangle}{\partial k} = \frac{1}{N} (p(\varphi) - \varphi).
\end{equation}
And for $\partial \sigma_{\Delta \varphi} / \partial k$ we have
\begin{equation}
\frac{\partial \sigma_{\Delta \varphi}}{\partial k} = \frac{\sigma_{\Delta \varphi}}{2k} - \frac{k (1-\varphi)\varphi}{\sigma_{\Delta \varphi} N^2(N-1)}.
\end{equation}
Then the equation for $k_{\rm min}$ is as follows
\begin{equation}
- \frac{\partial \langle \Delta \varphi \rangle}{\partial k} \Bigg|_{\varphi = \varphi_{\rm mid}} = \frac{\partial \sigma_{\Delta \varphi}}{\partial k} \Bigg|_{\varphi = \varphi_{\rm mid}}.
\label{kmin_eq}
\end{equation}
Since $H > 0$ and the intermediate point $\varphi_{\rm mid}$ is taken between the metastable state and the saddle point, the derivative $\partial \langle \Delta \varphi \rangle/ \partial k < 0$, so the minus sign was added in (\ref{kmin_eq}). The intermediate point lying between the metastable state $\varphi_{-}$ and the saddle point $\varphi_{0}$ can be defined as the point at which the restoring force (\ref{mean_dphi}) is maximal in magnitude:
\begin{equation}
\varphi_{\rm mid} = \arg \max_{\varphi \in [\varphi_{-}, \varphi_0]} | \langle \Delta \varphi \rangle |,
\end{equation}
This intermediate point $\varphi_{\rm mid}$ can be obtained by considering the corresponding derivative $\partial  \langle \Delta \varphi \rangle / \partial \varphi$.

\begin{figure}[ht]
	\center{\includegraphics[width=1\linewidth]{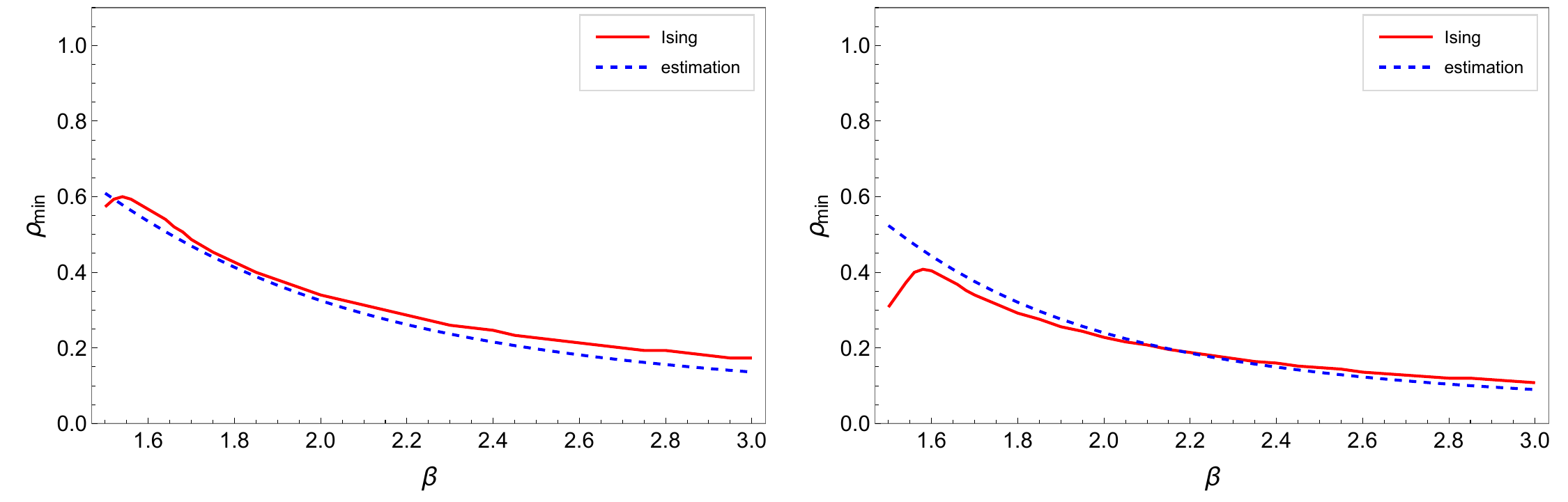}}
	\caption{\small Plots of $\rho_{\rm min}$ as a function of $\beta$ for $\gamma = 0.8$ and $N = 150$ (left), $N = 250$ (right). The exact and estimated minima $\rho_{\rm min}$ are plotted as solid and dashed lines, respectively.}
	\label{min_approx}
\end{figure}


Fig. \ref{min_approx} shows the plots of the minimum of the mean first-hitting time $\rho_{\rm min} = k_{\rm min} / N$ in the Ising game and the values of the minimum, which are estimated using the equation (\ref{kmin_eq}).

\begin{figure}[H]
	\center{\includegraphics[width=1\linewidth]{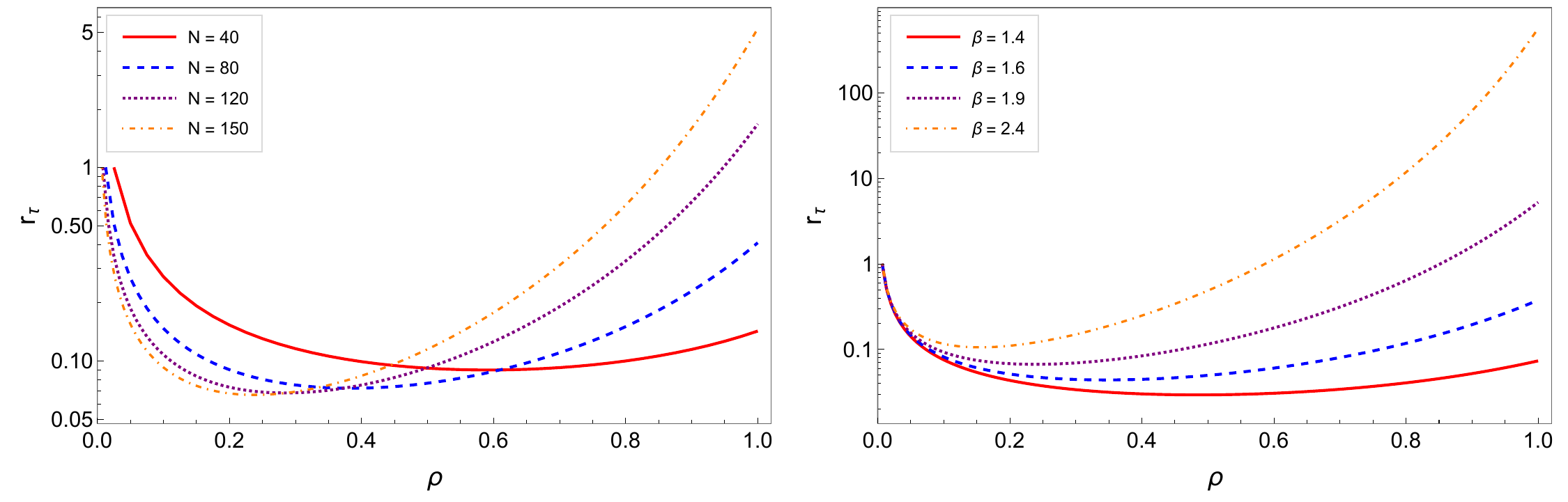}}
	\caption{\small Plots of $r_\tau (\beta, \gamma, \rho, N)$ as a function of $\rho$ for normally distributed noise with $\gamma = 0.8$ (1) for $N =40$ (solid line), $N =80$ (dashed line), $N = 120$ (dotted line), $N = 150$ (dash-dotted line) and $\beta = 1.9$ (left figure); (2) for $\beta = 1.4$ (solid line), $\beta = 1.6$ (dashed line), $\beta = 1.9$ (dotted line), $\beta = 2.4$ (dash-dotted line) and $N = 150$ (right figure).}
	\label{tau_th_normal}
\end{figure}

Previously, we considered the case of Gumbel noise, but a similar analysis can be performed for other types of stochastic disturbances, in particular, for normally distributed noise terms $\varepsilon$. In this case, the system still demonstrates the presence of low-temperature and high-temperature phases with the critical point $(\beta J)_{\text{crit}} = \sqrt{\pi} / 2$ \cite{Leonidov2024}. In the low-temperature phase, the value of $H^{*}$ can also be determined, but its finding requires numerical methods. Thus, the variable $\gamma$ can also be used to analyze the dynamics of the system. The Fig. \ref{tau_th_normal} shows the dependence of the normalized mean first-hitting time $r_\tau (\beta, \gamma, \rho, N)$ on the parameter $\rho$. The model parameters were chosen in such a way that the scale of the time characteristics remained comparable with the previously considered case of Gumbel noise. As can be seen from the plots, the qualitative nature of the dependence persists and corresponds to the general behavior of the system with Gumbel noise.

\subsection{Simulations}

Due to computational constraints and the exponentially long waiting time, it is infeasible to simulate the system until it exits the metastable state for arbitrarily selection of task parameters. Consequently, the parameters were chosen to ensure that simulations could be completed within a reasonable and computationally tractable timeframe. The primary numerical experiments were conducted with a system size of $N = 250$ (or $N = 300$) and $\gamma = 0.8$, i.e. an external field strength of $H = 0.8 H^{*}(\beta)$, where $H^{*}(\beta)$ denotes the critical field at inverse temperature $ \beta$. The evolution of the system begins from a metastable state and the game stops after the system hits the global minimum, in this case it is the right decision ($H > 0$).

\begin{figure}[ht]
	\center{\includegraphics[width=0.6\linewidth]{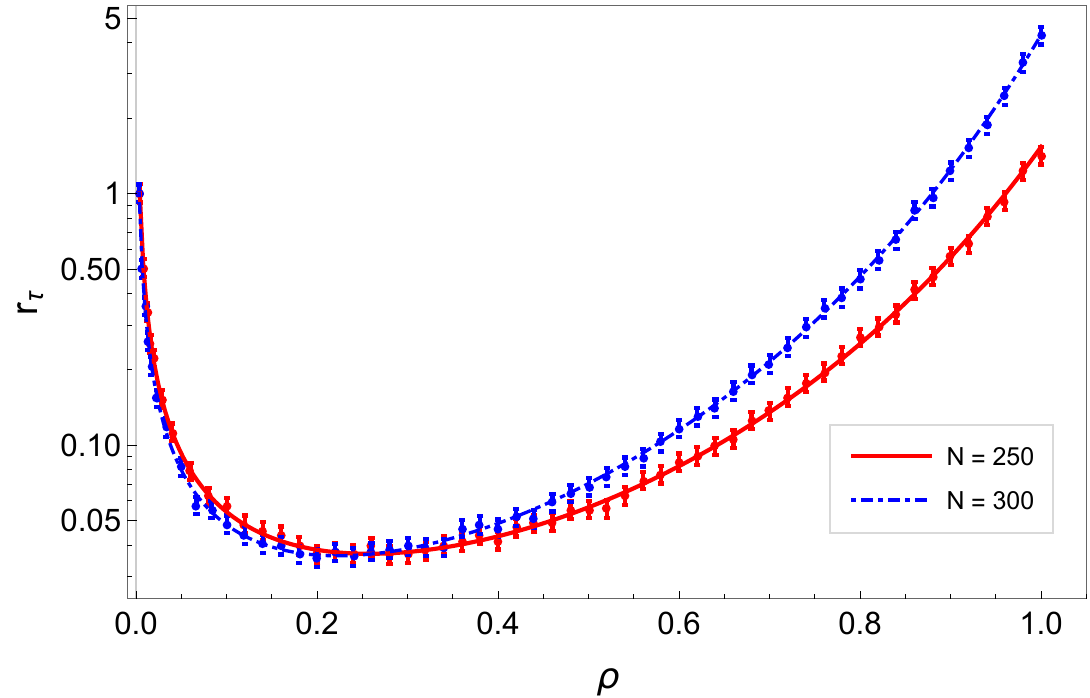}}
	\caption{\small Simulation data for \(r_\tau(\beta, H, \rho, N)\) as a function of \(\rho = k/N\)  for different values of $N$ with $\gamma = 0.8$ and $\beta = 1.9$. The lines show the values calculated from the theory of Markov chains.}
	\label{exp_and_th}
\end{figure}

For each fixed value of $k$, multiple independent realizations of the stochastic process were executed to gather statistical data on the transition time from the metastable state to the ground state. The transition times obtained from these simulations were then averaged arithmetically to estimate the mean first-hitting time.

\begin{figure}
	\center{\includegraphics[width=1\linewidth]{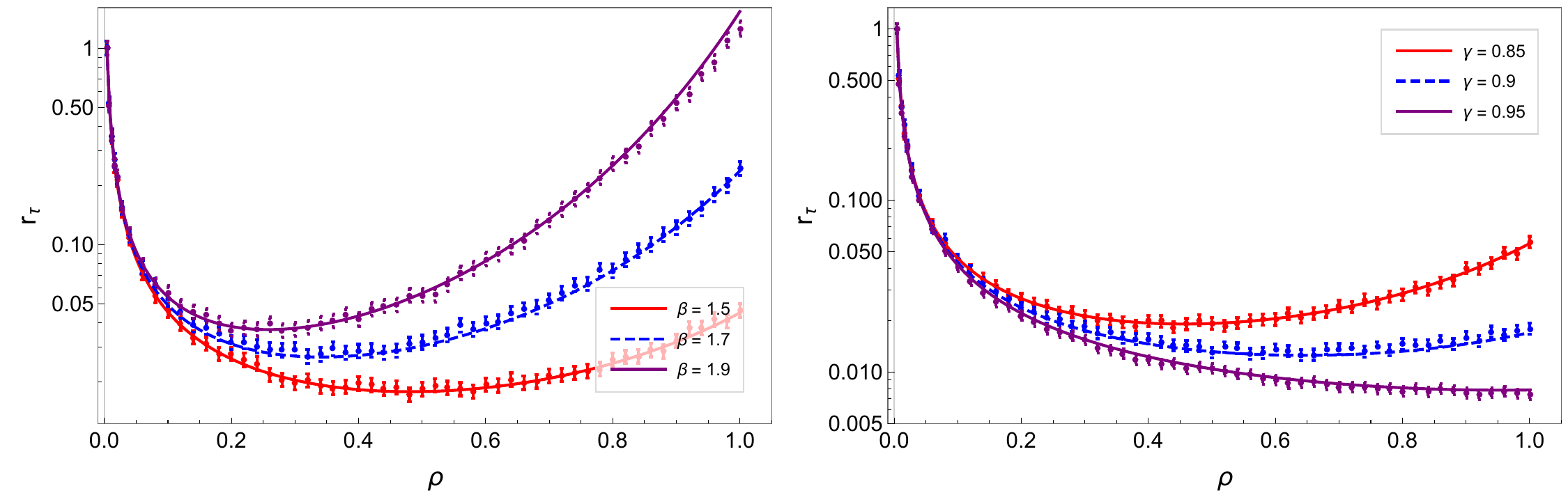}}
	\caption{\small Simulation data of \(r_\tau(\beta, H, \rho, N)\) as a function of \(\rho\) (1) for different values of  $\beta$ with \(\gamma = 0.8\) and $N = 250$ (left); (2) for different values of $\gamma$ with \(N = 250\) and $\beta = 1.7$ (right). The plots are shown on a logarithmic scale. The lines show the values calculated from the theory of Markov chains.}
	\label{exp_beta_h}
\end{figure}

Fig. \ref{exp_and_th} presents the experimental and theoretical mean first-hitting time as a function of $\rho$ for $N = 250$ and $N = 300$, including statistical error bars. At low values of $\rho$, there is a sharp drop in the transition time. With a further increase in $\rho$, there is an increase in the transition time. There is an excellent correspondence between theory and experiment.

In Fig. \ref{exp_beta_h} upper plots show the dependence of $r_\tau(\beta, H, \rho,N)$ on $\rho$ for different values of $\beta$. It can be seen that with an increase in $\beta$, the mean transition time grows significantly as $\rho\rightarrow 1$, as predicted by theory. At the bottom of \ref{exp_beta_h} are graphs of the dependence of $r_\tau(\beta, H, \rho,N)$ on $\rho$ for different values of $\gamma$. It can be seen that with an increase in $\gamma$, the growing dependence of $\langle\tau(\rho)\rangle$ disappears completely at $\rho\rightarrow 1$ and there is a change of two regimes, as was shown earlier from the results of calculating the mean first-hitting time in Markov chains.

\begin{figure}
	\center{\includegraphics[width=1\linewidth]{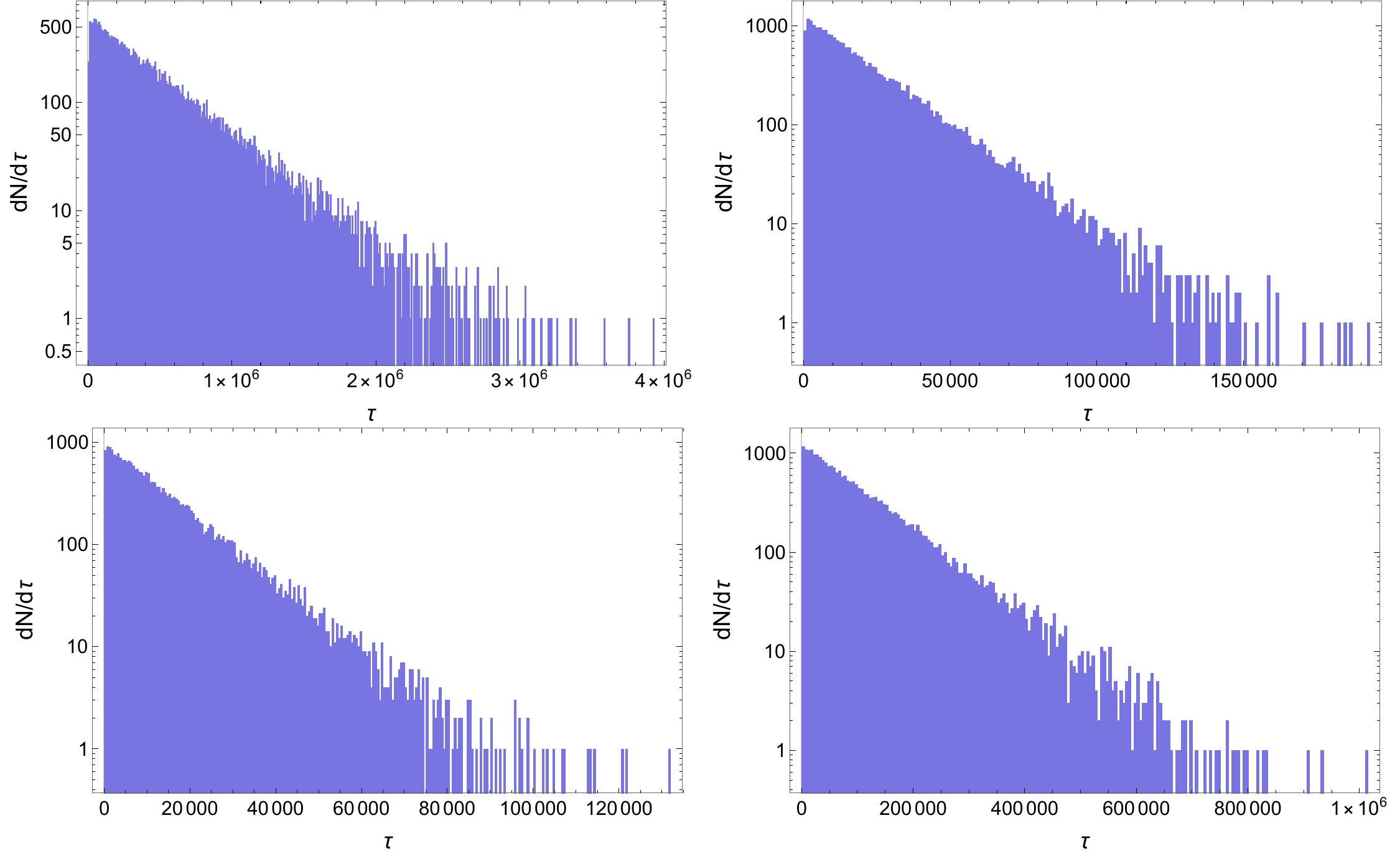}}
	\caption{\small Histograms of the first-hitting time distribution $\tau$ with \(N = 250, \beta = 1.7, \gamma = 0.8\) and \(k = 1\) (top left), \(k = 25\) (top right), \(k = 125\) (bottom left), \(k = 250\) (bottom right). The sample size is 25000.}
	\label{hist_N250B17_K_1_25_125_250}
\end{figure}

\begin{figure}[H]
	\center{\includegraphics[width=1\linewidth]{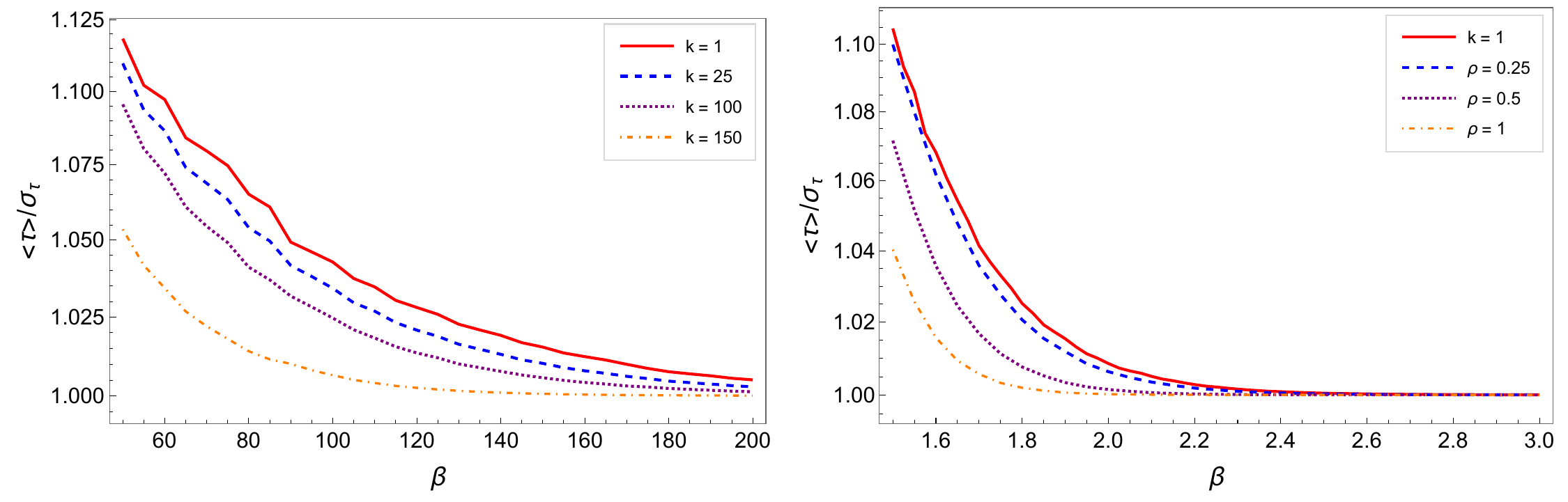}}
	\caption{\small The ratio  $\langle \tau \rangle / \sigma_\tau$  (1) as function of $\beta$ with \(N = 150, \gamma = 0.8\) and different values of $\rho$ ($k$); (2) as function of $N$ with \(\beta = 1.9, \gamma = 0.8\) and different values of $\rho$ ($k$).}
	\label{dist_ratio}
\end{figure}

In addition to analyzing the moments of the first-hitting time distribution, it is also of significant interest to study the corresponding histograms. Typical first-hitting time distributions were obtained and are presented on a logarithmic scale in Fig. \ref{hist_N250B17_K_1_25_125_250}. It can be observed that the shape of the distribution remains largely unchanged. Notably, the distribution tail decays exponentially, so the mean first-hitting time is well defined. It is known that for an exponential distribution, the first moment and standard deviation are the same. Figure \ref{dist_ratio} presents the ratio of the first moment $\langle \tau \rangle \equiv \langle \tau_{i^* j^*} \rangle$ to the standard deviation $\sigma_\tau$ for the first hitting time. As the parameters $N$ ($\beta$) increases, the ratio approaches unity, which may be indicative of the exponential nature of the underlying distribution.

\section*{Conclusion}

The paper studies the properties of $k$ -flip dynamics in the Ising-type systems of interacting binary units at an example of the noisy binary choice (Ising) game on complete graphs with uniform symmetric interactions. The discussion is framed in terms of Markov chain theory. The derivation of analytical expressions for the transition matrix $\Omega^{(k)}_{ij}$ and moments $\langle \Delta \varphi \rangle$ and second $\langle (\Delta \varphi )^2 \rangle$ is valid for arbitrary noise. The discussion of $k$ - dependence of the decay time of the metastable state is assuming the Gumbel noise distribution and, therefore, Boltzmannian expression for the single flip rate equivalent to the Glauber flip rate in statistical physics and neuron firing rate in the Little/Hopfield models in neurodynamics.

Let us list the main results obtained in the paper:
\begin{itemize}
\item analytical expression for the transition matrix $\Omega^{(k)}_{ij}$, equation \eqref{omegak} for arbitrary noise;
\item analytical expressions for the first $\langle \Delta \varphi \rangle$ and second $\langle (\Delta \varphi )^2 \rangle$ moments of the one-step increment $\Delta \varphi$ in equations \eqref{mean_dphi} and \eqref{second_monent} for arbitrary noise;
\item revelation of the U-shaped dependence of the decay time of the metastable state on $k$ and its qualitative explanation in terms of superimposed diffusive and restoring forces, paragraph 2.2.
\end{itemize}

One of the key questions in studying dynamical systems of the type considered in the present paper is the presence or absence of detailed balance. A discussion of this issue for the $k$-flip game on complete graphs with uniform symmetric interactions considered in this study is quite involved and will be published separately.

\printbibliography

@article{glauber1963time,
  title={Time-dependent statistics of the Ising model},
  author={Glauber, Roy J},
  journal={Journal of mathematical physics},
  volume={4},
  number={2},
  pages={294--307},
  year={1963},
  publisher={American Institute of Physics}
}

@article{derrida1989dynamical,
  title={Dynamical phase transitions in spin models and automata},
  author={Derrida, Bernard},
  journal={CEA-Conf-10034},
  year={1989},
  publisher={IOP}
}

@book{salinas2001introduction,
  title={Introduction to statistical physics},
  author={Salinas, Silvio},
  year={2001},
  address= {New {Y}ork},
  publisher={Springer New York, NY}
}

@article{cirillo2003metastability,
  title={Metastability for a Stochastic Dynamics with a Parallel Heat Bath Updating Rule},
  author={Cirillo, Emilio and Nardi, Francesca},
  journal={Journal of Statistical Physics},
  pages={183-217},
  year={2003},
  publisher={Springer}
}

@article{neri2009cavity,
  title={The cavity approach to parallel dynamics of Ising spins on a graph},
  author={Neri, Isaak and Bolle, Desire},
  journal={Journal of Statistical Mechanics: Theory and Experiment},
  volume={2009},
  number={08},
  pages={P08009},
  year={2009},
  publisher={IOP Publishing}
}

@article{avni2025nonreciprocal,
  title={Nonreciprocal Ising model},
  author={Avni, Yael and Fruchart, Michael and Martin, David and Seara, Daniel and Vitelli, Vincenzo},
  journal={Physical Review Letters},
  volume={134},
  number={11},
  pages={117103},
  year={2025},
  publisher={APS}
}

@article{garces2025phase,
  title={Phase transitions in single species Ising models with non-reciprocal couplings},
  author={Garces, Adria and Levis, Demian},
  journal={Journal of Statistical Mechanics: Theory and Experiment},
  pages={043205},
  year={2025},
  publisher={IOP}
}

@article{bagnoli2025metastability,
  title={Metastability in the diluted parallel Ising model},
  author={Bagnoli, Franco and Matteuzzi, Tommaso},
  journal={The European Physical Journal B},
  volume={98},
  pages={224},
  year={2025},
  publisher={Springer}
}

@article{little1974existence,
  title={The existence of persistent states in the brain},
  author={Little, William A},
  journal={Mathematical biosciences},
  volume={19},
  number={1-2},
  pages={101--120},
  year={1974},
  publisher={Elsevier}
}

@article{shaw1974persistent,
  title={Persistent states of neural networks and the random nature of synaptic transmission},
  author={Shaw, Gordon L and Vasudevan, R},
  journal={Mathematical Biosciences},
  volume={21},
  number={3-4},
  pages={207--218},
  year={1974},
  publisher={Elsevier}
}

@article{little1975statistical,
  title={A statistical theory of short and long term memory},
  author={Little, WA and Shaw, Gordon L},
  journal={Behavioral biology},
  volume={14},
  number={2},
  pages={115--133},
  year={1975},
  publisher={Elsevier}
}

@article{little1978analytic,
  title={Analytic study of the memory storage capacity of a neural network},
  author={Little, William A and Shaw, Gordon L},
  journal={Mathematical biosciences},
  volume={39},
  number={3-4},
  pages={281--290},
  year={1978},
  publisher={Elsevier}
}

@article{hopfield1982neural,
  title={Neural networks and physical systems with emergent collective computational abilities.},
  author={Hopfield, John J},
  journal={Proceedings of the national academy of sciences},
  volume={79},
  number={8},
  pages={2554--2558},
  year={1982}
}

@article{peretto1984collective,
  title={Collective Properties of Neural Networks: A Statistical Physics Approach.},
  author={Peretto, P},
  journal={Biological Cybernetics},
  volume={50},
  pages={51--62},
  year={1984}
}

@article{bastolla1998relaxation,
  title={Relaxation, closing probabilities and transition from oscillatory to chaotic attractors in asymmetric neural networks},
  author={Bagnoli, Franco and Matteuzzi, Tommaso},
  journal={The European Physical Journal B},
  volume={98},
  pages={224},
  year={2025},
  publisher={Springer}
}

@article{blume2003equilibrium,
  title={Equilibrium concepts for social interaction models},
  author={Blume, Lawrence and Durlauf, Steven},
  journal={International Game Theory Review},
  volume={5},
  number={03},
  pages={193--209},
  year={2003},
  publisher={World Scientific}
}

@article{bouchaud2013crises,
  title={Crises and collective socio-economic phenomena: simple models and challenges},
  author={Bouchaud, Jean-Philippe},
  journal={Journal of Statistical Physics},
  volume={151},
  number={3-4},
  pages={567--606},
  year={2013},
  publisher={Springer}
}

@article{antonov2021self,
  title={Self-excited Ising game},
  author={Antonov, A and Leonidov, A and Semenov, A},
  journal={Physica A: Statistical Mechanics and its Applications},
  volume={561},
  pages={125305},
  year={2021},
  publisher={Elsevier}
}

@article{leonidov2022strategic,
  title={Strategic stiffening/cooling in the Ising game},
  author={Leonidov, Andrey and Vasilyeva, Ekaterina},
  journal={Chaos, Solitons and Fractals},
  volume={100},
  pages={112279},
  year={2022},
  publisher={Elsevier}
}

@article{antonov2023transition,
  title={Transition between metastable equilibria: Applications to binary-choice games},
  author={Antonov, A and Leonidov, A and Semenov, A},
  journal={Physical Review E},
  volume={108},
  number={2},
  pages={024134},
  year={2023},
  publisher={APS}
}

@article{Leonidov2024,
title = {Ising game on graphs},
journal = {Chaos, Solitons \& Fractals},
volume = {180},
pages = {114540},
year = {2024},
issn = {0960-0779},
author = {Andrey Leonidov and Alexey Savvateev and Andrew G. Semenov}
}

@article{leonidov2024strategic,
  title={Strategic behaviour in the mean field  Ising game},
  author={Leonidov, A and Radionov, S and Vasilyeva, Ekaterina},
  journal={Chaos, Solitons and Fractals},
  volume={187},
  pages={115416},
  year={2024},
  publisher={Elsevier}
}

@article{garnier2024unlearnable,
  title={Unlearnable games and “satisficing” decisions: a simple model for a complex world},
  author={Garnier-Brun, Jerome and Benzaquen, Michael and Bouchaud, Jean-Philippe},
  journal={Physical Review X},
  volume={14},
  number={2},
  pages={021039},
  year={2024},
  publisher={APS}
}

@article{kasteleyn1969phase,
  title={Phase transitions in lattice systems with random local properties},
  author={Kasteleyn, PW and Fortuin, CM},
  journal={Journal of the Physical Society of Japan Supplement},
  volume={26},
  pages={11},
  year={1969}
}

@article{fortuin1972random,
  title={On the random-cluster model: I. Introduction and relation to other models},
  author={Fortuin, Cornelius Marius and Kasteleyn, Piet W},
  journal={Physica},
  volume={57},
  number={4},
  pages={536--564},
  year={1972},
  publisher={Elsevier}
}

@article{niedermayer1988general,
  title={General cluster updating method for Monte Carlo simulations},
  author={Niedermayer, Ferenc},
  journal={Physical review letters},
  volume={61},
  number={18},
  pages={2026},
  year={1988},
  publisher={APS}
}

@article{swendsen1987nonuniversal,
  title={Nonuniversal critical dynamics in Monte Carlo simulations},
  author={Swendsen, Robert H and Wang, Jian-Sheng},
  journal={Physical review letters},
  volume={58},
  number={2},
  pages={86},
  year={1987},
  publisher={APS}
}

@article{wolff1989collective,
  title={Collective Monte Carlo updating for spin systems},
  author={Wolff, Ulli},
  journal={Physical Review Letters},
  volume={62},
  number={4},
  pages={361},
  year={1989},
  publisher={APS}
}

@book{Kemeny1976,
  title={Finite Markov Chains},
  author={John G. Kemeny and J. Laurie Snell},
  year={1976},
  address= {New {Y}ork},
  publisher={Springer New York, NY},
}

@book{Grinstead1997,
  title={Introduction to Probability},
  author={Grinstead, C. M. and Snell, J. L.},
  year={1997},
  address= {Providence},
  publisher={American Mathematical Soc.},
}

@book{Risken1996,
  title={The Fokker-Planck Equation: Methods of Solution and Applications},
  author={Hannes Risken and Till Frank},
  year={1996},
  address = {Heidelberg},
  publisher={Springer},
}

\appendix

\section{Derivation of $\langle \Delta \varphi \rangle$}

For arbitrary \(k\), the mean $\langle\Delta \varphi \rangle$ is
\begin{align}
\langle\Delta \varphi \rangle &= \sum_{\{x\}} \frac{k!}{x_1!\ x_2!\ x_3!\ x_4!} \ p_+^{x_1 + x_3} (1-p_+)^{x_2 + x_4} \ \frac{N^{-}!}{(N^{-} - x_1 - x_2)!} \frac{N^{+}!}{(N^{+} - x_3 - x_4)!} \nonumber \\ &\times \frac{(N-k)!}{N!} \ \frac{1}{N} (x_1 - x_4),
\label{dm_true}
\end{align}
where $x_1$ is the number of outcomes when a player with \(s = 1\) is selected and changes strategy to \(s = -1\), i.e. $+1 \rightarrow -1$, then $x_2$ is the number of outcomes $+1 \rightarrow +1$ (i.e. does not change strategy), $x_3$ : $-1 \rightarrow -1$ and $x_4$ : $-1 \rightarrow +1$.

Let us fix \(x_1 + x_3\), \(x_2 + x_4\) as
\begin{equation}
x_1 + x_3 = n,
\end{equation}
and
\begin{equation}
x_2 + x_4 = k - n,
\end{equation}

Then (\ref{dm_true}) will be rewritten as
\begin{align}
&\sum_{x_1, x_4, n} \frac{k!}{x_1!\ (k - n - x_4)!\ (n - x_1)!\ x_4!}\ p_+^n (1-p_+)^{k-n} \nonumber \\ & \times \frac{N^{-}!}{(N^{-} - x_1 - k + n + x_4)!}\frac{N^{+}!}{(N^{+} - n + x_1 - x_4)!}  \frac{(N-k)!}{N!} \ \frac{1}{N} (x_1 - x_4).
\label{dm_false_2}
\end{align}
Let us first consider the sum over $x_1$ and $x_4$. The factor \(p_+^n (1-p_+)^{k-n} /N\) is irrelevant to this sum and can be omitted. We can explicitly separate the sums:
\begin{align}
&\sum_{x_1 = 0}^n \sum_{x_4 = 0}^{k-n} \frac{k!}{x_1!\ (k - n - x_4)!\ (n - x_1)!\ x_4!} \nonumber \\ & \times \frac{N^{-}!}{(N^{-} - x_1 - k + n + x_4)!} \frac{N^{+}!}{(N^{+} - n + x_1 - x_4)!}  \frac{(N-k)!}{N!}  (x_1 - x_4) \nonumber \\ & = \sum_{x_1 = 0}^n \sum_{x_4 = 0}^{k-n} \frac{k!}{x_1!\ (k - n - x_4)!\ (n - x_1)!\ x_4!} \nonumber \\ & \times \frac{(N-k)!}{(N^{-} - x_1 - k + n + x_4)!(N^{+} - n + x_1 - x_4)!} \frac{N^{+}!N^{-}!}{N!}  (x_1 - x_4).
\end{align}
One can rewrite everything in terms of binomial coefficients and also consider only the term multiplied by \(x_1\):
\begin{equation}
\sum_{x_1 = 0}^n \sum_{x_4 = 0}^{k-n} \binom{k}{n}\binom{n}{x_1}\binom{k-n}{x_4} \binom{N^{-} + N^{+} - k}{N^{-} - x_1 - k + n + x_4} \Bigg(\binom{N^{-} + N^{+}}{N^{-}} \Bigg)^{-1} \ x_1.
\end{equation}

For binomial coefficients we use the Vandermonde convolution
\begin{equation}
\sum_k  \binom{r}{m + k} \binom{s}{n - k} = \binom{r + s}{m + n}.
\end{equation}
Then we have
\begin{align}
&\sum_{x_4 = 0}^{k-n} \binom{k-n}{x_4} \binom{N^{-} + N^{+} - k}{N^{-} - x_1 - k + n + x_4} \nonumber \\ & = \sum_{x_4 = 0}^{k-n} \binom{k-n}{k - n - x_4} \binom{N^{-} + N^{+} - k}{N^{-} - x_1 - k + n + x_4} = \binom{N^{-} + N^{+} - n}{N^{-} - x_1}.
\end{align}
It follows from the property of binomial coefficients that
\begin{equation}
\binom{n}{x_1}\ x_1 = n\binom{n - 1}{x_1 - 1}.
\end{equation}
Taking into account the above expressions , we obtain
\begin{equation}
\sum_{x_1 = 0}^n \binom{k}{n}n\binom{n - 1}{x_1 - 1} \binom{N^{-} + N^{+} - n}{N^{-} - x_1} \Bigg(\binom{N^{-} + N^{+}}{N^{-}} \Bigg)^{-1}.
\label{v1}
\end{equation}
In (\ref{v1}) we see a similar Vandermonde convolution:
\begin{align}
&\sum_{x_1 = 0}^{n}\binom{n - 1}{x_1 - 1} \binom{N^{-} + N^{+} - n}{N^{-} - x_1} \nonumber \\ & = \binom{N^{-} + N^{+} - 1}{N^{-} - 1}.
\end{align}
Then for the sum with $x_1$ we have
\begin{align}
&\binom{k}{n}n \binom{N^{-} + N^{+} - 1}{N^{-} - 1} \Bigg(\binom{N^{-} + N^{+}}{N^{-}} \Bigg)^{-1} = \binom{k}{n}n \frac{(N^{-} + N^{+} - 1)!}{N^{+}!(N^{-} - 1)!} \frac{N^{+}! N^{-}!}{(N^{-}  + N^{+})!}  \nonumber \\ & = \binom{k}{n}n \frac{N^{-}}{N^{-} + N^{+}} =  \binom{k}{n}n \frac{N^{-}}{N}.
\label{true_ans}
\end{align}

Similarly, we can obtain for the term with \(x_4\). Thus, we obtain
\begin{align}
&\sum_{x_1 = 0}^n \sum_{x_4 = 0}^{k-n} \binom{k}{n}\binom{n}{x_1}\binom{k-n}{x_4} \binom{N^{-} + N^{+} - k}{N^{-} - x_1 - k + n + x_4} \Bigg(\binom{N^{-} + N^{+}}{N^{-}} \Bigg)^{-1} \ x_4 \nonumber \\ & = \binom{k}{n} (k-n) \frac{N^{+}}{N}.
\label{dm_x4}
\end{align}

The final expression for $\langle\Delta \varphi \rangle$ is given by:
\begin{equation}
\langle\Delta \varphi \rangle = \sum_{n}   \binom{k}{n} \Bigg[n \frac{N^{-}}{N} - (k-n) \frac{N^{+}}{N} \Bigg] \ p_+^n (1-p_+)^{k-n}  \frac{1}{N} = \frac{k}{N} (p_+ - \varphi) .
\label{dm_true_3}
\end{equation}

\section{Derivation of $\langle (\Delta \varphi)^2 \rangle$}

The expression for $\langle ( \Delta \varphi )^2 \rangle$ has the form:
\begin{align}
\langle (\Delta \varphi)^2 \rangle &=\sum_{x_1, x_4, n} \frac{k!}{x_1!\ (k - n - x_4)!\ (n - x_1)!\ x_4!}  p_+^n (1-p_+)^{k-n}  \label{dm_true_2} \\ &\times \frac{N^{-}!}{(N^{-} - x_1 - k + n + x_4)!} \frac{N^{+}!}{(N^{+} - n + x_1 - x_4)!} \frac{(N-k)!}{N!} \frac{1}{N^2} (x_1 - x_4)^2 \nonumber.
\end{align}
One can expand $ (x_1 - x_4)^2 = (x_1^2 -2x_1 x_4 + x_4^2)$. Let's factor out the sum over \(n\) and consider the sum for the first term \(x_1^2\):
\begin{align}
& \sum_{x_1 = 0}^n \sum_{x_4 = 0}^{k-n} \binom{k}{n}\binom{n}{x_1}\binom{k-n}{x_4} \binom{N^{-} + N^{+} - k}{N^{-} - x_1 - k + n + x_4} \nonumber \\ &\times \Bigg(\binom{N^{-} + N^{+}}{N^{-}} \Bigg)^{-1} \ \frac{1}{N^2} x_1^2.
\label{x2_term}
\end{align}
Using the Vandermonde convolution, we can write
\begin{align}
&\sum_{x_4 = 0}^{k-n} \binom{k-n}{x_4} \binom{N^{-} + N^{+} - k}{N^{-} - x_1 - k + n + x_4} \sum_{x_4 = 0}^{k-n} \binom{k-n}{k - n - x_4} \binom{N^{-} + N^{+} - k}{N^{-} - x_1 - k + n + x_4} \nonumber \\ & = \binom{N^{-} + N^{+} - n}{N^{-} - x_1}.
\end{align}

Then (\ref{x2_term}) reads
\begin{equation}
\sum_{x_1 = 0}^n \binom{k}{n}\binom{n}{x_1}\binom{N^{-} + N^{+} - n}{N^{-} - x_1} \Bigg(\binom{N^{-} + N^{+}}{N^{-}} \Bigg)^{-1} \ \frac{1}{N^2} x_1^2.
\label{x1_2}
\end{equation}

Make the following transformation
\begin{align}
&\binom{n}{x_1} x_1^2 = n\binom{n - 1}{x_1 - 1} x_1 = n\binom{n - 1}{x_1 - 1} (x_1 - 1 + 1) \nonumber \\ & = n(n-1) \binom{n - 2}{x_1 - 2} + n\binom{n - 1}{x_1 - 1}.
\end{align}

For the first term we use the following convolution
\begin{align}
&\sum_{x_1 = 0}^{n}\binom{n - 2}{x_1 - 2} \binom{N^{-} + N^{+} - n}{N^{-} - x_1} = \binom{N^{-} + N^{+} - 2}{N^{-} - 2}.
\end{align}
And for the second term we have
\begin{align}
&\sum_{x_1 = 0}^{n}\binom{n - 1}{x_1 - 1} \binom{N^{-} + N^{+} - n}{N^{-} - x_1} = \binom{N^{-} + N^{+} - 1}{N^{-} - 1}.
\end{align}

In this case, putting everything together, for (\ref{x1_2}) we will have
\begin{align}
&\Bigg[ n(n-1) \binom{N^{-} + N^{+} - 2}{N^{-} - 2} + n \binom{N^{-} + N^{+} - 1}{N^{-} - 1} \Bigg] \binom{k}{n} \Bigg(\binom{N^{-} + N^{+}}{N^{-}} \Bigg)^{-1}\frac{1}{N^2} \nonumber \\ &= n(n-1) \binom{k}{n} \frac{(N^{-} + N^{+} - 2)!}{N^{+}!(N^{-} - 2)!} \frac{N^{+}! N^{-}!}{(N^{-}  + N^{+})!}\frac{1}{N} +\binom{k}{n}n \frac{N^{-}}{N^{-} + N^{+}}\frac{1}{N^2} \nonumber \\ &=  n(n-1) \binom{k}{n} \frac{N^{-}(N^{-} -1)}{(N^{-}  + N^{+})(N^{-}  + N^{+} - 1)}\frac{1}{N^2} +\binom{k}{n}n \frac{N^{-}}{N^{-} + N^{+}}\frac{1}{N^2} \nonumber \\ &= \binom{k}{n}n \frac{(1 - \varphi)}{N^2} \Bigg( (n-1) \frac{N(1 - \varphi) - 1}{N - 1} + 1 \Bigg),
\label{x1_2_ans}
\end{align}
where we moved to the variables \(\varphi = N^{+} / N, \ N\).

The result of a similar calculation for the term \(x_4^2\) is obtained by making the substitution in (\ref{x1_2_ans}) \(n \rightarrow k - n\) (without touching the binomial coefficient) and \(\varphi \rightarrow 1 - \varphi\):
\begin{align}
\binom{k}{n} (k-n)  \frac{\varphi}{N^2} \Bigg( (k-n-1) \frac{N\varphi - 1}{N - 1} + 1 \Bigg),
\label{x4_2_ans}
\end{align}

For the term \(- 2x_1 x_4\) we will have
\begin{equation}
\sum_{x_1 = 0}^n \sum_{x_4 = 0}^{k-n} \binom{k}{n}\binom{n}{x_1}\binom{k-n}{x_4} \binom{N^{-} + N^{+} - k}{N^{-} - x_1 - k + n + x_4} \Bigg(\binom{N^{-} + N^{+}}{N^{-}} \Bigg)^{-1} \ \frac{1}{N^2} (-2x_1 x_4).
\label{x1x4}
\end{equation}

There are several transformations:
\begin{equation}
\binom{n}{x_1} x_1 = n\binom{n - 1}{x_1 - 1}
\label{vv1}
\end{equation}
and
\begin{equation}
\binom{k-n}{x_4} x_4 = (k-n)\binom{k - n - 1}{x_4 - 1}.
\label{vv2}
\end{equation}

In this case, the Vandermonde convolution gives
\begin{equation}
\sum_{x_4 = 0}^{k-n} \binom{k-n - 1}{x_4 - 1} \binom{N^{-} + N^{+} - k}{N^{-} - x_1 - k + n + x_4} = \binom{N^{-} + N^{+} - n - 1}{N^{-} - x_1}.
\label{vv3}
\end{equation}

Let's substitute (\ref{vv1}), (\ref{vv2}), (\ref{vv3}) into (\ref{x1x4}):
\begin{equation}
-\sum_{x_1 = 0}^n 2 \binom{k}{n} n (k-n) \binom{n - 1}{x_1 - 1} \binom{N^{-} + N^{+} - n - 1}{N^{-} - x_1} \Bigg(\binom{N^{-} + N^{+}}{N^{-}} \Bigg)^{-1} \frac{1}{N^2}.
\end{equation}
The next convolution gives
\begin{align}
&\sum_{x_1 = 0}^{n} \binom{n - 1}{x_1 - 1} \binom{N^{-} + N^{+} - n - 1}{N^{-} - x_1} = \binom{N^{-} + N^{+} - 2}{N^{-} - 1} 
\end{align}

Then we get
\begin{align}
& -\binom{k}{n} n (k-n) \binom{N^{-} + N^{+} - 2}{N^{-} - 1} \Bigg(\binom{N^{-} + N^{+}}{N^{-}} \Bigg)^{-1}\frac{1}{N^2} \nonumber \\ &= -2 n(k-n) \binom{k}{n} \frac{(N^{-} + N^{+} - 2)!}{(N^{+} - 1)!(N^{-} - 1)!} \frac{N^{+}! N^{-}!}{(N^{-}  + N^{+})!}\frac{1}{N^2} \nonumber \\ &= -2 n(k-n) \binom{k}{n} \varphi (1 - \varphi) \frac{1}{N(N - 1)}.
\label{x1x4_ans}
\end{align}

Summing expressions (\ref{x1_2_ans}, \ref{x4_2_ans}, \ref{x1x4_ans}) gives
\begin{align}
\langle (\Delta \varphi)^2 \rangle &=\sum_{n = 0}^k p_+^n (1-p_+)^{k-n}\frac{1}{N^2} \Bigg[\binom{k}{n}n (1 - \varphi) \Bigg( (n-1) \frac{N(1 - \varphi) - 1}{N - 1} + 1 \Bigg)  \nonumber \\ &- 2 n(k-n) \binom{k}{n} \varphi (1 - \varphi) \frac{N}{N - 1}  \nonumber \\ & +  \binom{k}{n} (k-n)  \varphi \Bigg( (k-n-1) \frac{N\varphi - 1}{N - 1} + 1 \Bigg) \Bigg] .
\end{align}

After some simplifications, we will have
\begin{equation}
\langle (\Delta \varphi)^2 \rangle = \sum_{n = 0}^k \binom{k}{n} p_+^n (1-p_+)^{k-n} \frac{1}{N^2} \Bigg[n^2 -2kn\varphi + k\varphi\frac{-k + N + (k - 1)N\varphi}{N-1}\Bigg]  \nonumber.
\end{equation}

After transformations
\begin{equation}
\binom{k}{n} n^2 = k(k-1) \binom{k - 2}{n - 2} + k\binom{k - 1}{n - 1}
\end{equation}
and
\begin{equation}
\binom{k}{n} n = k\binom{k - 1}{n - 1},
\end{equation}
we get the final answer:
\begin{equation}
\langle (\Delta \varphi)^2 \rangle = \frac{k}{N^2} \Bigg( p_+^2 (k-1) + p_+ (1 - 2k\varphi) + \varphi\frac{-k + N + (k - 1)N\varphi}{N-1} \Bigg).
\end{equation}

\section{Derivation of $\langle (\tau_{ij})^2 \rangle$}

For the second moment $\langle (\tau_{ij})^2\rangle$ we have
\begin{equation}
\langle (\tau_{ij})^2 \rangle = 1 \cdot \Omega_{ij} + \sum_{n \neq j} \Omega_{in}  \langle (\tau_{nj} + 1)^2 \rangle = 1 + \sum_{n \neq j} \Omega_{in}  \langle (\tau_{nj})^2 \rangle + 2 \sum_{n \neq j} \Omega_{in}  \langle (\tau_{nj}) \rangle.
\label{var_tau}
\end{equation}

Denote the value of $\langle (\tau_{ij})^2\rangle = T_{ij}^{(2)}$, then in matrix form for (\ref{var_tau}) we get
\begin{equation}
T^{(2)} = E + \Omega T^{(2)}  - \Omega \diag T^{(2)} + 2 \Omega T - 2 \Omega\diag T.
\label{t2_matrix}
\end{equation}

To go further, it is necessary to understand the properties of the fundamental martix $Z$. The following relations are valid for the matrix $Z$:
\begin{itemize}
\item $Z^{-1} \Pi = (I - \Omega + \Pi) \Pi = \Pi \ \Rightarrow \ Z \Pi = \Pi Z = \Pi.$
\item  $Z^{-1} E = (I - \Omega + \Pi) E = E \ \Rightarrow \ Z E = E Z = E.$
\item $Z Z^{-1} = Z (I - \Omega + \Pi) = I \ \Rightarrow \ Z (I - \Omega) = I - \Pi.$
\item  $Z \Omega = \Omega Z,$
\end{itemize}
where $E$ is a matrix with all its matrix elements equal to one. In addition, for matrices $\Pi$ and $D = {\rm diag} (1/\pi)$, the following holds: $\Pi D = E$.

It is known from expression (\ref{m_ij}) that $\diag T = D$. Let us obtain an expression for $\diag T^{(2)}$. To do this, multiply the original equation (\ref{t2_matrix}) from the left by $\Pi$ and the the diagonal, then
\begin{equation}
\Pi T^{(2)} = E + \Pi T^{(2)}  - \Pi \diag T^{(2)} + 2 \Pi T - 2 \Pi D.
\end{equation}
That gives us the expression for $\Pi \diag T^{(2)}$:
\begin{equation}
\Pi \diag T^{(2)} = E + 2 \Pi T - 2 \Pi D.
\end{equation}
Note that $T = (E \diag Z - Z + I)D$ and $\Pi D = E$, then we have
\begin{equation}
\Pi \diag T^{(2)} = - E + 2 \Pi (E \diag Z - Z + I) D = - E + 2 E D \diag Z.
\end{equation}
If we expand the indicies, we get
\begin{equation}
\sum_k \Pi_{ik} T^{(2)}_{kj} \delta_{kj} = - 1 + 2 Z_{jj} D_{jj},
\end{equation}
where $\delta_{kj}$ is the Kronecker symbol. Using $\Pi_{ik} = \pi_k, D_{jj} = 1/\pi_j$ we have
\begin{equation}
\pi_j T^{(2)}_{jj} = - 1 + 2 \frac{Z_{jj}}{\pi_j}.
\end{equation}
And we get the final expression for the matrix $\diag T^{(2)}$:
\begin{equation}
\diag T^{(2)}= - D + 2 E D^2 \diag Z.
\label{diag_t2}
\end{equation}

Now the equation (\ref{t2_matrix}) reads
\begin{equation}
(I - \Omega) T^{(2)} = E - \Omega( - D + 2 E D^2 \diag Z) + 2 \Omega(I + E \diag Z - Z) D - 2\Omega D.
\label{diag_t2_2}
\end{equation}
Multiply the equation (\ref{diag_t2_2}) by $Z$ on the left, then we get
\begin{equation}
Z (I - \Omega) T^{(2)} = E - Z\Omega( - D + 2 E D^2 \diag Z)+ 2 Z\Omega(I + E \diag Z - Z) D - 2Z\Omega D.
\end{equation}
Using the properties of the fundamental matrix $Z(I - \Omega) = (I - \Pi)$ and $Z\Omega = (Z + \Pi - I)$ one can obtain
\begin{align}
(I - \Pi) T^{(2)}&= E - (Z + \Pi - I)( - D + 2 E D^2 \diag Z) \nonumber \\ &+ 2 (Z + \Pi - I)(I + E \diag Z - Z) D - 2(Z + \Pi - I)\Omega D.
\end{align}
After expanding all the brackets and simplifying, we have
\begin{equation}
(I - \Pi) T^{(2)} = 3ZD - 2 Z D^2 \diag Z - 2 Z^2 D - D + 2 D^2 \diag Z.
\label{diag_t2_3}
\end{equation}

Since
\begin{equation}
\big( \Pi T^{(2)} \big)_{ij} = \sum_k \Pi_{ik} T^{(2)}_{kj} = \sum_k \pi_k T^{(2)}_{kj} = \big( \Pi T^{(2)} \big)_{jj}
\end{equation}
is valid, only the diagonal elements of the matrix $\Pi T^{(2)}$ need to be determined. One can take the diagonal of expression (\ref{diag_t2_3}):
\begin{align}
& - \diag \big( \Pi T^{(2)} \big)  - D + 2 D^2 \diag Z \nonumber \\ &= 3D \diag Z \ 2 D^2 (\diag Z)^2 - 2 D \diag (Z^2) - D + 2 D^2 \diag Z \\
& \diag \big( \Pi T^{(2)} \big) = - 3D \diag Z + 2 D^2 (\diag Z)^2 + 2 D \diag (Z^2).
\end{align}
Therefore, for the matrix $\Pi T^{(2)}$ one can write
\begin{equation}
\Pi T^{(2)} = -3ED \diag Z + 2 E D^2 (\diag Z)^2 + 2 E D \diag (Z^2).
\label{pi_t2}
\end{equation}

Substituting (\ref{pi_t2}) into (\ref{diag_t2_3}), we get
\begin{align}
T^{(2)} &= 3ED \diag Z - 2 E D^2 (\diag Z)^2 - 2 E D \diag (Z^2) \nonumber \\ &+ 3ZD - 2 Z D^2 \diag Z - 2 Z^2 D - D + 2 D^2 \diag Z.
\end{align}
After some simplifications, using $T = (E\diag Z - Z + I)D$, we obtain the final answer for the matrix $T^{(2)}$:
\begin{equation}
T^{(2)} = 2 T D \diag Z - 3T+ 2 (I + E\diag(Z^2) - Z^2)D.
\label{t2_final}
\end{equation}

\end{document}